\documentclass [11pt, letterpaper]{article}
\pdfoutput=1
\usepackage[english]{babel}
\usepackage{fullpage}
\usepackage[normalem]{ulem}
\usepackage{amsmath}
\usepackage{amssymb}
\usepackage{graphicx}
\usepackage{mathrsfs}
\usepackage{wrapfig}
\usepackage{boxedminipage}
\usepackage{epsfig}
\usepackage{setspace}
\usepackage{subfigure}
\usepackage[unicode=true]{hyperref}
\usepackage[usenames]{color}
\newcommand{\reef}[1]{(\ref{#1})}
\title{\bf Entanglement Entropy\\ \bf of\\ \bf Magnetic Electron Stars}
\bigskip
\author{Tameem Albash, Clifford V. Johnson, Scott MacDonald}
\date{}

\begin{document}
\onehalfspacing

\bigskip
\maketitle
\centerline{\it Department of Physics and Astronomy}
\centerline{\it University of Southern California}
\centerline{\it Los Angeles, CA 90089-0484, U.S.A.}
\bigskip

\centerline{\small \tt albash, johnson1, smacdona, [at] usc.edu}

\bigskip

\begin{abstract}
We study the behavior of the entanglement entropy in $(2+1)$--dimensional strongly coupled theories via the AdS/CFT correspondence.  We consider theories at a finite charge density with a magnetic field, with their holographic dual being  Einstein-Maxwell-Dilaton theory in four dimensional anti--de Sitter gravity.   Restricting to black hole and electron star solutions at zero temperature in the presence of a background magnetic field, we compute their holographic entanglement entropy using the Ryu-Takayanagi prescription for both strip and disk geometries.  In the case of the electric or magnetic zero temperature black holes, we are able to confirm that the entanglement entropy is invariant under electric-magnetic duality.  In the case of the electron star with a finite magnetic field, for the strip geometry, we find a discontinuity in the first derivative of the entanglement entropy as the strip width is increased.
\end{abstract}

\newpage
\section{Introduction}
The AdS/CFT correspondence \cite{Maldacena:1997re,Witten:1998qj,Gubser:1998bc,Aharony:1999t} has become a prevalent new theoretical tool for understanding strongly--coupled, non--gravitational physics.  Its influence has extended beyond the realm of strict high energy physics, like understanding the quark--gluon plasma \cite{Kovtun:2004de,Iqbal:2008by}, into condensed matter systems.   Of interest are strongly--coupled charged fermionic systems, where it is hoped the low--energy physics can be studied via the correspondence (see refs.~\cite{Witten:1998zw,McGreevy:2009xe,Hartnoll:2009sz,Hartnoll:2011fn}). In particular, progress has been made in understanding Fermi Liquids (FL), non--Fermi Liquids (NFL) and Fractionalized Fermi Liquids (FL*), phases of compressible metallic states of quantum matter \cite{Ogawa:2011bz,Huijse:2011ef}\footnote{See refs. \cite{Sachdev:2010um,Huijse:2011hp} for explanations of FL, NFL and FL* phases, where we have used their abbreviations for the names of the phases.}.  

These different phases with finite charge density can be characterized in terms of their violation/agreement with the Luttinger relation, which relates the total charge density $Q$ to the volumes enclosed by the Fermi surfaces at zero temperature.  Compressible states of matter with finite charge density dual to charged black holes \cite{Lee:2008xf,Liu:2009dm,Cubrovic:2009ye,Faulkner:2009wj} violate the Luttinger relation. However, holographic duals to metallic states, such as the electron star solutions of refs.~\cite{Hartnoll:2010xj,Hartnoll:2011dm,Hartnoll:2010gu}, do satisfy the Luttinger relation \cite{Iqbal:2011in,Sachdev:2011ze}.  The essential idea \cite{Huijse:2011ef} is that when the field theory charge density $Q$ is dual to gauge-invariant fermions -- ``mesinos" -- in the gravity bulk, their Fermi surfaces do satisfy the Luttinger relation and are said to have \textit{visible} Fermi surfaces. When the charge density is instead sourced by a charged horizon, then the Fermi surfaces are said to be \textit{hidden}, leading to a violation of the Luttinger relation and thus describe ``fractionalized" charged degrees of freedom. In the case when there are both gauge--invariant fermions and a charged horizon in the bulk, like the solutions in ref.~\cite{Hartnoll:2011pp}, then the phase is considered to be ``partially fractionalized."

Among the interesting questions posed about NFL and FL* phases is the issue of how to characterize the presence of Fermi surfaces when the conventional methods from field theory are not easily applicable. It was shown in ref.~\cite{Ogawa:2011bz} that such phases in the field theory have a logarithmic violation of the area law for the entanglement entropy, and this sets a strict criterion for gravitational models to be considered dual to such phases.  The prescription of Ryu and Takayanagi \cite{Ryu:2006bv} for the calculation of the holographic entanglement entropy allows us to directly address this question. 

 Results of holographic computations suggest that the mesonic phases are dual to FLs, whereas the fractionalized phases are dual to NFLs and partially fractionalized to FL*s \cite{Hartnoll:2011pp}. In particular, it was found in refs. \cite{Ogawa:2011bz,Huijse:2011ef} that the hidden, gauge--charged Fermi surfaces in the bulk do lead to a logarithmic violation of the area law when the metric is of hyperscaling--violating form with particular values of the dynamical critical exponent and the hyperscaling violating exponent. In the case of the electron star, the logarithmic violation depends only on the charge sourced by the horizon and not that coming from the star itself \cite{Huijse:2011ef}.

 Here we extend existing work in the literature by computing the entanglement entropy of solutions to a $(3+1)$--dimensional Einstein--Maxwell--Dilaton (EMD) theory in the bulk that are not of hyperscaling--violating type and additionally have an external, constant background magnetic field turned on, including the magnetic electron star solutions found in ref.~\cite{Albash:2012ht}.

 Magnetic fields are also important probes of the physics of transport (especially for $(2+1)$--dimensional systems, to which our gravity solutions are dual), and  our work is intended to be part of a program of trying to understand, using such external fields, possible new ways to classify  characteristic behaviours in the phases we capture. In this paper we will focus on studies of the behaviour of the entanglement entropy  in the presence of the magnetic field. Since, as stated above (see also further discussion below) our geometries are not of hyperscaling--violating type,  we do not expect  the characteristic logarithmic violations of the area law, and so we seek to explore and exhibit  the entanglement entropy's behavior in the various regimes to which we have access.

 Our work studies gravitational backgrounds not previously studied in the literature; in the notation of ref.~\cite{Huijse:2011ef}, their $\alpha$ and $\beta$, which determine the dilaton's potential and coupling to the Maxwell sector, in our cases are equal: $\alpha=\beta$. In terms of the dynamical critical exponent $z$, this corresponds to $z\rightarrow\infty$ and a hyperscaling violating exponent $\theta=2$. These values are incompatible with the form of the hyperscaling--violating metric, and in particular do not satisfy the requirements of ref.~\cite{Huijse:2011ef} that $\beta \leq \frac{1}{3}\alpha$ and $\theta = 1$. 
 
 In ref.~\cite{Kundu:2012jn}, the authors consider an EMD theory using a potential for the dilaton of the form $V(\Phi)=-\vert V_{0} \vert \text{exp}\left(2\delta\Phi\right)$ and coupling to the Maxwell sector $Z(\Phi)=\text{exp}\left(2\alpha\Phi\right)$. They study the entanglement entropy of purely electric, purely magnetic (argued for via electromagnetic duality), and dyonic solutions of this system, adding the magnetic field as a small perturbation, $B\ll\mu^2$, in their dyonic system. (Here, $\mu$ is the chemical potential.) In their notation, our solutions correspond to the choice $\alpha=\delta$, which is a case they do not consider in their work\footnote{In ref.~\cite{Iizuka:2011hg}, the authors consider a fermionic two--point function for $\alpha = \delta$ and find that it exhibits non--Fermi liquid behavior.}. In particular, they start with a hyperscaling--violating metric for the purely electric solution parameterized by their $(\alpha,\delta)$, but in the case $\alpha=\delta$, the dynamical exponent and the hyperscaling violating exponent both diverge, positively and negatively, respectively. Thus, our EMD solutions lie along a line in their phase diagram for which they do not explore the entanglement entropy. In addition, our dyonic solutions have $Q=B$.
 
We present our work as follows.  In Section \ref{EEReview} we begin with a review of the holographic entanglement entropy, but present the detailed derivations of the formulae we display in the Appendix. In Section \ref{GravityBackgrounds} we present our gravity backgrounds, although for more information on their derivation and the magnetic electron star solutions we refer to ref. \cite{Albash:2012ht}. Section \ref{PED} shows the results for the strip and disk (see Section \ref{EEReview} for a definition of these) entanglement entropy for the purely electric dilaton black holes, and Section \ref{PMD} for the purely magnetic dilaton black holes. Section \ref{EMDual} studies the behavior of the strip entanglement entropy under electromagnetic duality for the purely electric and purely magnetic dilaton black holes. In particular, it is shown how the entanglement entropy is invariant as long as one is careful to take into consideration the relative positions of the physical horizons. The entanglement entropy for the strip and disk for the dilaton--dyon black hole is presented in Section \ref{DD}, and then in Section \ref{MES} we consider the mesonic phase of the magnetic electron star solutions of ref. \cite{Albash:2012ht}. We end in Section \ref{Conclusion} with our conclusions and a discussion of future work\footnote{While this work was in preparation, two papers appeared on the arXiv, refs. \cite{Carney:2015dra,Puletti:2015gwa} that construct magnetic electron stars that are different from the ones presented in ref.~\cite{Albash:2012ht}. In this new work, there is no dilaton present in the theory and their horizons have finite temperature. They do not consider the entanglement entropy of their solutions, but it would be interesting to compute the entanglement entropy for their solutions and compare them to the results in this paper.}.

\section{Review of the Holographic Entanglement Entropy} \label{EEReview}
The holographic entanglement entropy of Ryu and Takayanagi \cite{Ryu:2006bv} computes the entropy of entanglement $S_{A}$ between two subsystems $A$ and $B=\bar{A}$ (the complement of $A$). The prescription involves finding the minimal area surface that extends into the AdS gravitational bulk, whose boundary at conformal infinity is that of subsystem $A$. That is, if $\gamma_{A}$ is the minimal surface in the $(d+2)$--dimensional bulk such that $\partial\gamma_{A}=A$ at the $(d+1)$--dimensional UV boundary, then
\begin{equation}
S_{A}=\frac{\text{Area}(\gamma_{A})}{4G_{N}^{(d+2)}} \ ,
\label{eqn:EEdef}
\end{equation}
where the surface $\gamma_{A}$ has co--dimension $2$ and $G_{N}^{(d+2)}$ is Newton's constant in $(d+2)$--dimensions. To compute the area, ones takes the bulk metric $G_{\mu\nu}$ and integrates its pull-back $H_{\mu\nu}$ onto $\gamma_{A}$ for a constant time slice $t=t_{0}$:
\begin{equation}
\text{Area}(\gamma_{A})=\int_{\gamma_{A}}{d^{d}x\sqrt{H}} \ .
\label{eqn:areaformula}
\end{equation}
Here $x$ represents the coordinates on $\gamma_{A}$, which are generally given as the embedding of the surface, and $H$ is the determinant of $H_{\mu\nu}$.

The above prescription was described for $(d+2)$--dimensional AdS spacetime, but it applies more generally. In particular, there may be a non-zero and dynamical dilaton $\Phi$  in the higher ten--dimensional string theory. When this is the case, eqn.~\reef{eqn:areaformula} still holds so long as the metric is written in Einstein frame \cite{Ryu:2006ef}.

\begin{figure}[h!]
\centering
\subfigure[Strip]{\includegraphics[width = 3in]{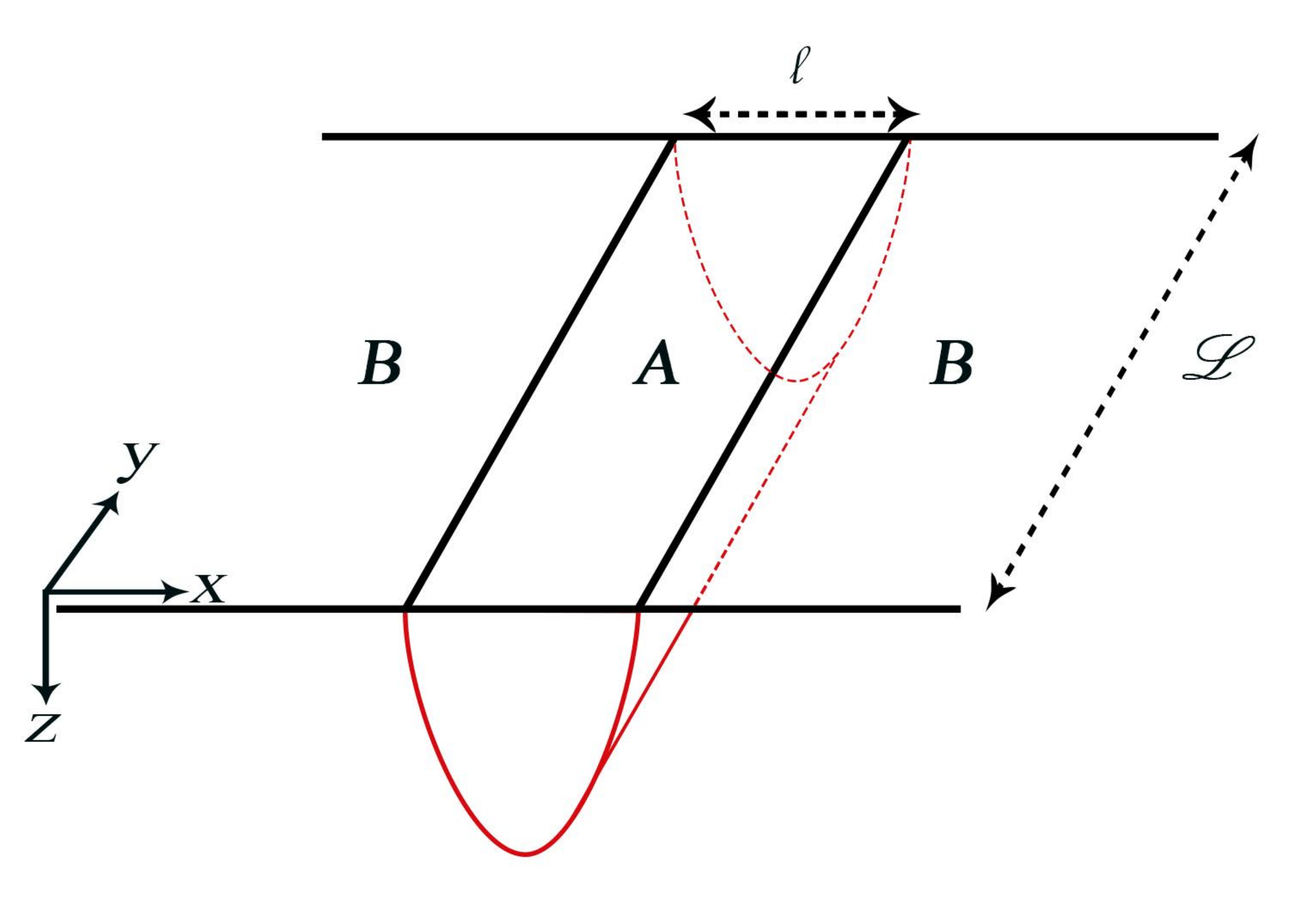}}
\subfigure[Disk]{\includegraphics[width = 3in]{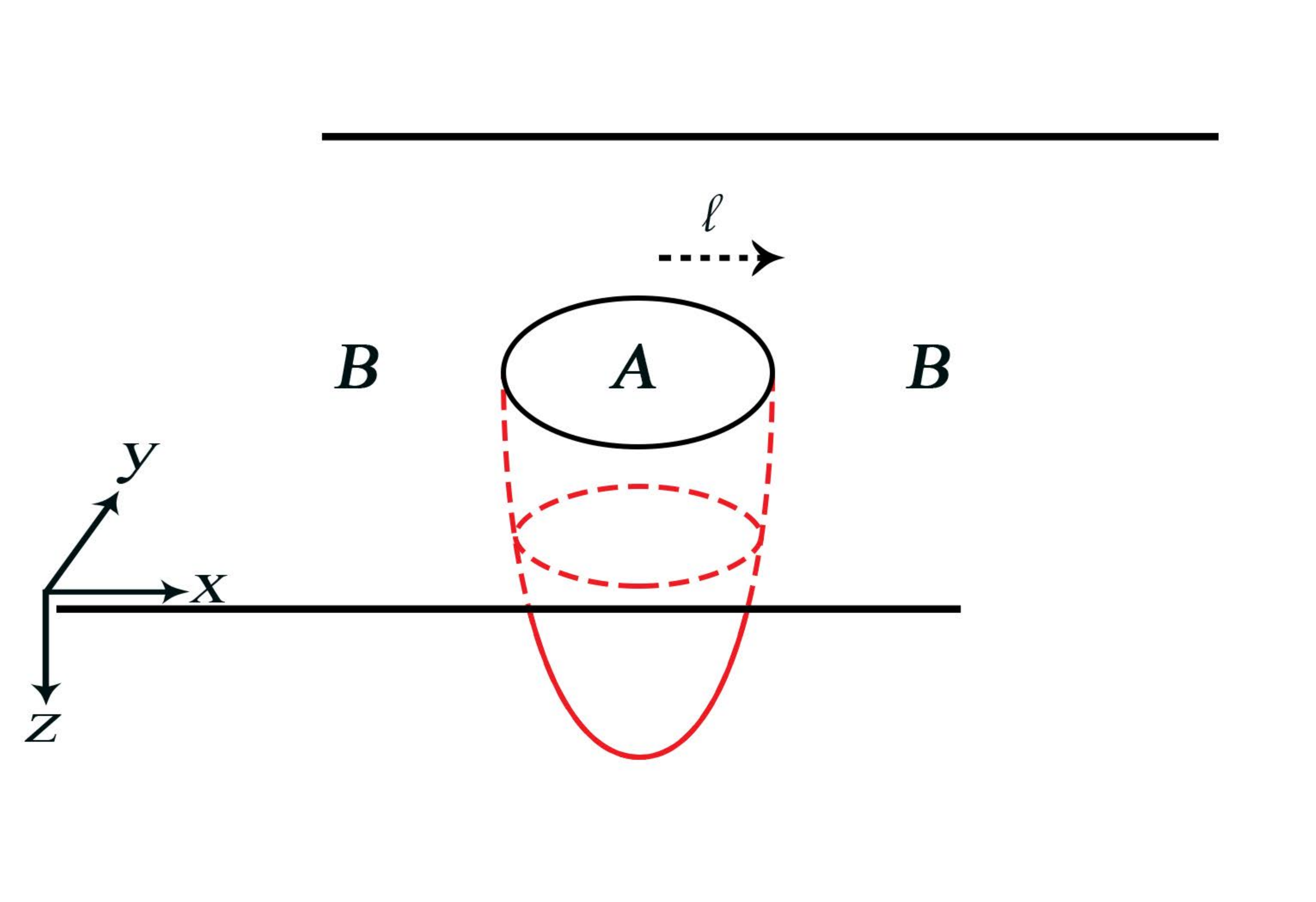}}
\caption{A schematic representation of the two regions, $A$ (with complement $B$), on the boundary (with coordinates $\{x,y\}$) whose holographic entanglement entropy we consider, where the holographic direction is $z$. The minimal surfaces hang down into the bulk. (a) The strip, whose finite entanglement entropy we denote $s_{S}$, with finite width $\ell$. (b) The disk, whose finite entanglement entropy we denote $s_{D}$, with radius $\ell$.}
\label{fig:EEbdygeometries}
\end{figure}

 In what follows, we will work with a four--dimensional bulk (so that $d=2$ in eqn.~\reef{eqn:EEdef} and hence will drop any sub-- or superscripts denoting the dimension) with Newton's constant in eqn.~\reef{eqn:EEdef} given by dimensional reduction in the usual way, and the regions we consider in the boundary are that of the strip and the disk as seen in Figure~\ref{fig:EEbdygeometries}; we shall denote their entanglement entropies as~$S_{S}$ and~$S_{D}$, respectively. In particular, we will be interested in the finite part of the entanglement entropy, $s_{S}$ and $s_{D}$, defined by
\begin{align}
4 G_{N} S_{S}&=2 L^{2} \mathscr{L}\left(s_{S}+\frac{1}{\epsilon}\right) \ , \label{eqn:EEstrip}\\
4 G_{N} S_{D}&=2 \pi L^{2}\left(s_{D}+\frac{\ell}{\epsilon}\right) \ . \label{eqn:EEdisk}
\end{align}
Here $\mathscr{L}$ denotes the strip geometry's ``infinite" side length, $L$ is the AdS length scale, $\epsilon$ is a chosen~UV cutoff, and $\ell$ is the radius of the disk. The added term on the right hand side of eqns.~\reef{eqn:EEstrip} and \reef{eqn:EEdisk} removes the leading divergence of the entanglement entropy, and we note that these expressions have been defined for dimensionless coordinates so that $\mathscr{L}$, $\epsilon$, $\ell$, and the $s_{S/D}$ are all dimensionless.

\indent For reference, we now present generic formulae for the finite entanglement entropy for the strip and the disk for a form  of the bulk metric that all of our geometries will have:
\begin{equation}
ds^2=-g_{tt}(z)dt^2+g_{zz}(z)dz^2+g_{xx}(z)d\vec{x}^2 \ ,
\end{equation}
where we have taken the UV to be at $z=0$ and the horizon  to be at some $z=z_{H}$ that we rescale to be at $z=1$. Writing the $\mathbb{R}^2$ part of the metric as $dx^2+dy^2$ and having the finite length of the strip run from $-\ell/2 \leq x\leq \ell/2$, the minimal surface will be symmetric about $y=0$. We denote the~$z$ value of the turning point of the surface as~$z_{T}$. Let the surface have coordinates $(x,y)$, the same as the $\mathbb{R}^2$ coordinates, with embedding $z=z(x)$.

 In the Appendix we derive the formulae for the entanglement entropy for the strip and the disk geometries. We note that in the case of the strip, the second order problem of finding the minimal surface $z(x)$ can be reduced to a first order problem via an integral of motion (see the Appendix for details), which allows the integrals to be written in the form shown below; in addition, it grants more control on the numerics of our work in the following Sections. For the strip, the result is\footnote{Note that here we have not yet subtracted the leading AdS divergence nor pulled out a factor of the AdS radius~$L$, so that these are not quite the expressions for $s_{S}$ and $s_{D}$.}
\begin{equation}
4G_{N}S_{S}=2\mathscr{L}\int_{\epsilon}^{z_{T}}{dz \frac{g_{xx}(z)^2}{\sqrt{\frac{g_{xx}(z)}{g_{zz}(z)}\left(g_{xx}(z)^2-g_{xx}(z_{T})^2\right)}}} \ ,
\label{eqn:EEstripintegral}
\end{equation}
where we have introduced the UV cutoff $\epsilon$ in the lower limit of the integral. The length of the strip, $\ell=2\int_{0}^{\ell/2}{dx}$, as a function of the turning point is given by
\begin{equation}
\frac{\ell}{2}=\int_{0}^{z_{T}}{dz\frac{g_{xx}(z_{T})}{\sqrt{\frac{g_{xx}(z)}{g_{zz}(z)}\left(g_{xx}(z)^2-g_{xx}(z_{T})^2\right)}}} \ .
\label{eqn:EEstriplength}
\end{equation}

 In the case of the disk geometry, the full second order problem must be solved, so the formula for the entanglement entropy is simply given by the pull--back
\begin{equation}
4G_{N}S_{D}=2\pi\int_{0}^{\ell}{dr\, r g_{xx}(z(r))\sqrt{1+\frac{g_{zz}(z(r))}{g_{xx}(z(r))}\left(\frac{dz}{dr}\right)^2}} \ ,
\label{eqn:EEdiskpullback}
\end{equation}
where $0\leq r\leq \ell$ is the radial variable for polar coordinates of the $\mathbb{R}^2$, i.e., $ds^2_{\mathbb{R}^2} = dr^2+r^2d\theta^2$. In this case, we must explicitly solve for the minimal surface $z(r)$ and then input that into eqn.~\reef{eqn:EEdiskpullback} to find the entanglement entropy. Because of this more numerically intensive approach, we will observe some increased numerical variance for the disk results relative to the strip in the Sections that follow.

\section{Gravity Backgrounds} \label{GravityBackgrounds}
Our main goal will be to study the entanglement entropy for the magnetic electron star solution found in ref.~\cite{Albash:2012ht}, but to understand that result in a larger context, we will consider other related backgrounds as well. Here we present the common aspects of these solutions, giving the precise forms of the metrics below when we compute their entanglement entropy.

We will begin by considering zero temperature black hole solutions of a four--dimensional Einstein--Maxwell--Dilaton system with action
\begin{equation}
S_{\text{EMD}}=\int d^{4}x\sqrt{-G}\left[\frac{1}{2\kappa^{2}}\left(R-2\partial_{\mu}{\Phi}\partial^{\mu}{\Phi}-V(\Phi)\right)-\frac{Z(\Phi)}{4e^{2}}F_{\mu\nu}F^{\mu\nu}\right] \ .
\label{eqn:EMDaction}
\end{equation}
Here $\kappa^2 = 8\pi G_{N}$ is the gravitational coupling and $e$ is the Maxwell coupling. We take the potential for the dilaton and its coupling to the Maxwell field to be
\begin{equation}
V(\Phi)=-\frac{6}{L^2}\cosh{\left(2\Phi/\sqrt{3}\right)},\qquad Z(\Phi)=e^{2\Phi/\sqrt{3}} \ .
\end{equation}

We will consider solutions that have either just electric or just magnetic charge, as well as solutions with both. Our ansatz for the metric and Maxwell fields with a constant magnetic field~$B$ turned on is
\begin{align}
ds^2 &=  \frac{L^2}{z^2}\left(-F(z)dt^2+G(z)dz^2+A(z)d\vec{x}^2\right) \ , \label{eqn:metric} \\
F_{zt} &= \frac{eL}{\kappa z_{H}}h'(z), \qquad F_{xy} = \frac{eL}{\kappa z_{H}^2}B \ . \label{eqn:maxwell}
\end{align}
The horizon is at $z_{H}$ and the UV at $z=0$. When we wish to consider cases where there is no electric or no magnetic field, we will find it consistent to simply set $h(z)\equiv 0$ or $B\equiv 0$, respectively. The coordinates are dimensionful; however, the fields $h$ and $B$ are dimensionless. We will require our solutions to asymptote to $\text{AdS}_{4}$ in the UV, which dictates the leading behavior of the fields to be
\begin{subequations}
\begin{align}
F(z)&=1+\dots \ ,\qquad G(z)=1+\dots \ ,\qquad A(z)=1+\dots \ , \label{eqn:UVmetricbehavior}\\
h(z) &= \mu - \frac{Q}{z_{H}} z+\dots \ ,\qquad \Phi(z)=\frac{\phi_{1}}{z_{H}} z+\frac{\phi_{2}}{z_{H}^2} z^2+\dots \ .
\end{align}
\end{subequations}
The $\textit{physical}$ quantities in the dual field theory are the chemical potential $\mu_{P}=\frac{eL}{\kappa z_{H}}\mu$, the charge density $Q_{P}=\frac{L}{e\kappa z_{H}^2}Q$, the magnetic field $B_{P}=\frac{eL}{\kappa z_{H}^2}B$, the source for the operator dual to the dilaton $\phi_{1P}=\frac{L}{\kappa z_{H}}\phi_{1}$, and the vacuum expectation value (vev) of the operator dual to the dilaton $\phi_{2P}=\frac{L}{\kappa z_{H}^2}\phi_{2}$. It is convenient to work with dimensionless quantities, and so, by taking ratios with respect to $\mu_{P}$, we can characterize the dual field theory by the following dimensionless ratios:
\begin{equation}
\frac{B_{P}}{\mu_{P}^2}=\frac{\kappa}{eL}\frac{B}{\mu^2} \ ,\qquad \frac{Q_{P}}{\mu_{P}^2}=\frac{\kappa}{eL}\frac{Q}{\mu^2} \ ,\qquad \frac{\phi_{1P}}{\mu_{P}}=\frac{1}{e}\frac{\phi_{1}}{\mu} \ ,\qquad \frac{\phi_{2P}}{\mu_{P}^2}=\frac{\kappa}{e^2 L}\frac{\phi_{2}}{\mu^2}.
\label{eqn:dimensionlessratios}
\end{equation}
In order not to specify values for $e$, $L$, and $\kappa$, we will give the values of $B/\mu^2$, $Q/\mu^2$, etc. in what follows. We also work in dimensionless coordinates defined by replacing $(z, \ t, \ \vec{x})$ in the expressions above by $(z_{H}z,\ z_{H}t, \ z_{H}\vec{x})$, such that the horizon position is at $z=1$.

 We now summarize our work in ref. \cite{Albash:2012ht} on how to introduce the charged star, and refer to that reference for the full details. We take as our action
\begin{equation}
S=S_{\text{EMD}}+S_{\text{fluid}},
\label{eqn:EMDFluidaction}
\end{equation}
where the fluid action comes from the Lagrangian \cite{deRitis:1985uw,Bombelli:1990ze,Brown:1992kc}
\begin{equation}
\mathcal{L}_{\text{fluid}}=\sqrt{-G}\left(-\rho(\sigma)+\sigma u^{\mu}\left(\partial_{\mu}\phi+A_{\mu}+\alpha\partial_{\mu}\beta\right)+\lambda\left(u^{\mu}u_{\mu}+1\right)\right).
\end{equation}
Here $\phi$ is a Clebsch potential variable, $(\alpha,\beta)$ are potential variables, and $\lambda$ is a Lagrange multiplier. The energy density of the fluid, $\rho$, and the charge density, $\sigma$, were found in ref.~\cite{Albash:2012ht} to be that of a free fermion with mass $\tilde{m}=\kappa m/e$:
\begin{align}
\tilde{\sigma} &= eL^2\kappa\sigma= \frac{1}{3}\tilde{\beta}\left(\tilde{\mu}-\tilde{m}^2\right)^{3/2},\\
\tilde{\rho}&=L^2\kappa^2\rho= \frac{1}{8}\tilde{\beta}\left(\tilde{\mu}\sqrt{\tilde{\mu}^2-\tilde{m}^2}\left(2\tilde{\mu}^2-\tilde{m}^2\right)+\tilde{m}^4\ln{\left(\frac{\tilde{m}}{\tilde{\mu}+\sqrt{\tilde{\mu}^2-\tilde{m}^2}}\right)}\right),
\end{align}
where $\tilde{\mu}(z)=z h(z)/\sqrt{F(z)}$, $\tilde{\beta}=e^{4}L^{2}\beta/\kappa^2\sim\mathcal{O}(1)$ is a constant of proportionality, and a tilde means that it is a dimensionless quantity. The free fermion result is valid so long as $2\tilde{q}B \ll 1$, where $\tilde{q}=\frac{\kappa}{e^{2}L}\frac{2e}{a(z)}$ and $a(z)=A(z)/z^{2}$. To be in a regime where we can use classical gravity we must have $\kappa/L\ll 1$, and since $\tilde{\beta}\sim\mathcal{O}(1)$ it follows that $e^{2}\sim\kappa/L\ll 1$. Thus the free fermion result is valid when $\frac{4eB}{a(z)}\ll 1$. In the solutions that follow it is assumed that we adjust the value of $e$ so as to maintain this constraint. Our ansatz for the fluid is
\begin{equation}
u_{t}=-\sqrt{-G_{tt}},\qquad A_{y}=\frac{eL}{\kappa}By,\qquad \phi=\frac{eL}{\kappa}Bxy, \qquad \alpha=\frac{eL}{\kappa}By,\qquad \beta=x.
\end{equation}

With this ansatz the equations of motion from the action in eqn.~\reef{eqn:EMDFluidaction} become, after some work, 
\begin{subequations}
\begin{align}
& \tilde{P}'(z)+\frac{f'(z)}{2f(z)}\left( \tilde{P}(z)+\tilde{\rho}(z)\right)-\tilde{\sigma}(z)\frac{h'(z)}{\sqrt{f(z)}}=0 \ , \label{eqn:pressure}\\
& a''(z)-a'(z)\left( \frac{g'(z)}{2g(z)}+\frac{f'(z)}{2f(z)}+\frac{a'(z)}{2a(z)}\right) + a(z)\left( g(z)\left( \tilde{P}(z)+\tilde{\rho}(z)\right) + 2\Phi'(z)^2\right) = 0 \ ,\\
& \frac{f''(z)}{f(z)}-\frac{f'(z)}{f(z)}\left( \frac{g'(z)}{2g(z)} +\frac{f'(z)}{2f(z)} -\frac{2a'(z)}{a(z)}\right) \nonumber \\ 
& \hspace{3cm}+ \left(\frac{a'(z)^2}{2a(z)^2}-2\Phi'(z)^2-g(z)\left(5\tilde{P}(z)+\tilde{\rho}(z)-2\tilde{V}(\Phi)\right)\right)=0 \ ,\\
& \Phi'(z)^2+g(z)\left(-\frac{Z(\Phi)B^2}{2a(z)^2}+\left(\tilde{P}(z)-\frac{1}{2}\tilde{V}(\Phi)\right)\right) - \frac{a'(z)^2}{4a(z)^2}-\frac{a'(z)f'(z)}{2a(z)f(z)}-\frac{Z(\Phi)h'(z)^2}{2f(z)}=0 \ ,\\
& h''(z)-h'(z)\left(\frac{g'(z)}{2g(z)}+\frac{f'(z)}{2f(z)}-\frac{a'(z)}{a(z)}-\frac{Z'(\Phi)\Phi'(z)}{Z(\Phi)}\right)-\frac{\sqrt{f(z)}g(z)}{Z(\Phi)}\tilde{\sigma}(z)=0 \ ,\\
& \Phi''(z)+\Phi'(z)\left(\frac{f'(z)}{2f(z)}-\frac{g'(z)}{2g(z)}+\frac{a'(z)}{a(z)}\right) - \frac{Z'(\Phi)}{4}\left(\frac{g(z)B^2}{a(z)^2}-\frac{h'(z)^2}{f(z)}\right)-\frac{g(z)\tilde{V}'(\Phi)}{4}=0 \ .
\end{align}
\label{eqn:eom}
\end{subequations}
In these equations, we have written $f(z)=F(z)/z^2$, $g(z)=G(z)/z^2$, $a(z)=A(z)/z^2$, and $\tilde{V}(\Phi)=L^2 V(\Phi)$. Here the fluid pressure $\tilde{P}(z)$ is given by the thermodynamic relation~$\tilde{P}(z) = -\tilde{\rho}(z)+\tilde{\mu}(z)\tilde{\sigma}(z)$ so that eqn.~\reef{eqn:pressure} is automatically satisfied. As a check, setting the fluid fields to zero reduces eqn.~\reef{eqn:eom} to those coming from just the Einsten--Maxwell--Dilaton system in eqn.~\reef{eqn:EMDaction},  so when we consider cases without the star we will find solutions to eqn.~\reef{eqn:eom} with the fluid fields turned off. We still require our star backgrounds to be asymptotically $\text{AdS}_{4}$, and so the UV dual field theory is again characterized by the dimensionless ratios in eqn.~\reef{eqn:dimensionlessratios}.

\begin{figure}[h!]
\centering
\includegraphics[width=0.4\textwidth]{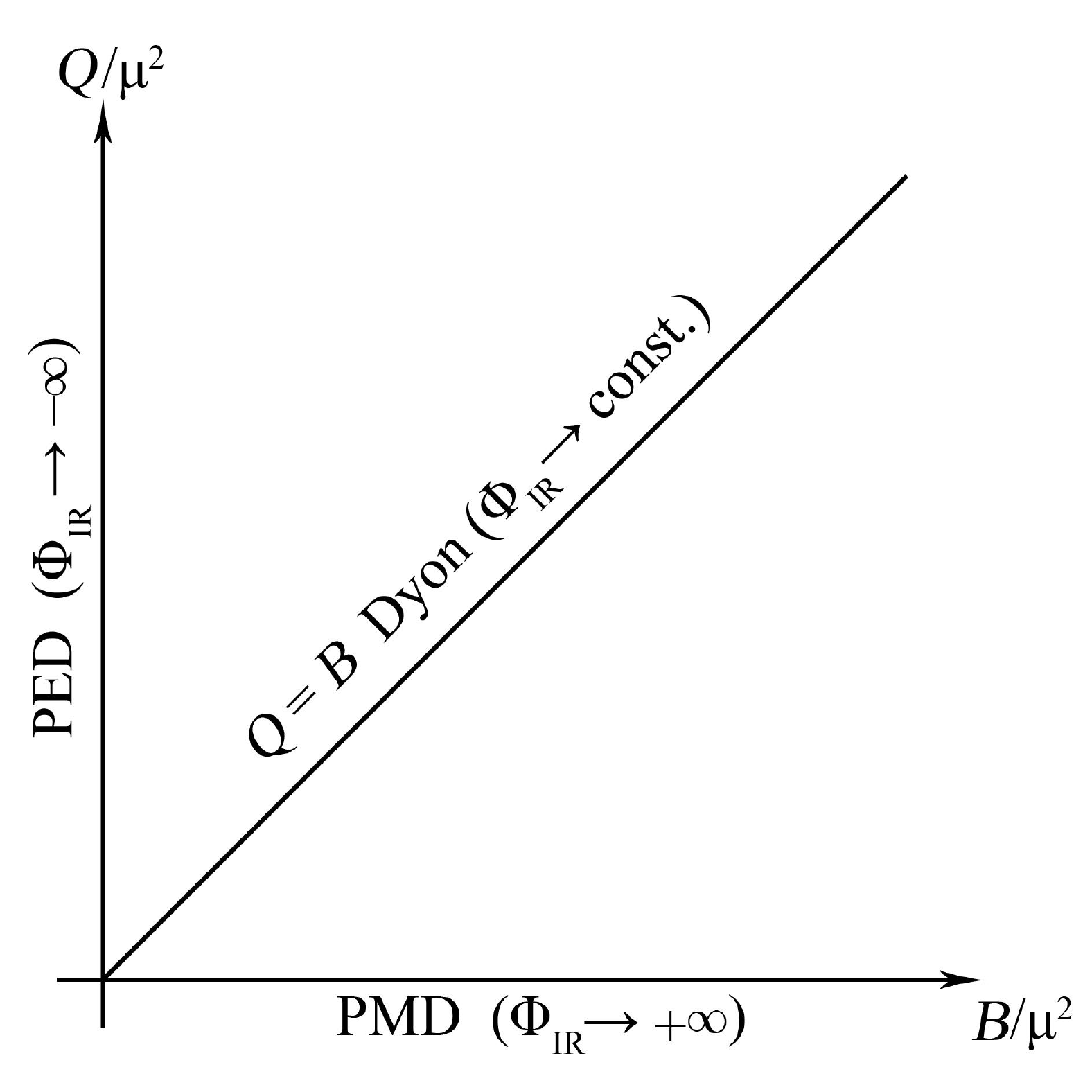}
\caption{Schematic phase diagram of the various black hole solutions considered in this paper. The vertical axis is the purely electric dilaton black hole (PED) with no magnetic charge, the horizontal axis is the purely magnetic dilaton black hole (PMD) with no electric charge, and the line is the dilaton--dyon black hole, with magnetic and electric charge satisfying $Q=B$. The IR behavior of the dilaton is indicated for each phase, in coordinates where the horizon is located at infinity.}
\label{fig:phasediagram}
\end{figure}

 It was found in ref.~\cite{Albash:2012ht} that the solutions of these equations fall into three broad phases governed by the IR behavior of the dilaton. In the case with no star, the dilaton can diverge logarithmically either positively or negatively in the IR, giving rise to a purely electric horizon or a purely magnetic horizon. If it tends to a finite value, then it is a dyonic solution. Figure~\ref{fig:phasediagram} schematically illustrates this phase structure. With the star present, the same classification scheme can be used and we can label the solutions: a ``mesonic" phase, where all the electric charge is sourced by the star; a ``partially fractionalized" phase, where a fraction of the charge is sourced by the star and the rest by the horizon; and a ``fully fractionalized" phase, where all the charge is sourced by the horizon. 
 
 In this paper, we will study the entanglement entropy of the following phases: no star purely electric horizon, no star purely magnetic horizon, no star dilaton--dyon, and star mesonic phase. These solutions were found in ref.~\cite{Albash:2012ht} as asymptotic expansions in the IR and were integrated numerically out to the UV, matching onto $\text{AdS}_{4}$. We display the relevant expansions as they are needed in the following sections.

\section{Entanglement Entropy of the Purely Electric Dilaton Black Hole} \label{PED}
We first begin with the purely electric dilaton black hole (PED) with no star. An analytic form of this solution was written down in ref.~\cite{Gubser:2009qt} (where it arises as the three--equal--charge dilatonic black hole in four dimensions) at a single point in the phase diagram; below, we indicate which point this is in terms of parameters. Since our star backgrounds are constructed numerically, as a check of our numerical procedure, we construct a numerical PED below and compare the results with the analytic solution of ref.~\cite{Gubser:2009qt}.

 Our IR ($z=1$) series expansion for the metric \reef{eqn:metric} is
\begin{subequations}
\begin{align}
F(z) &= (1-z)^{3/2}\left(f_{0}\left(1+\sum_{n=1}^{\infty}f_{n}(1-z)^{n}\right)+\delta f (1-z)^{b}\right) \ ,\\
G(z) &=  (1-z)^{-3/2}\left(\sum_{n=0}^{\infty}g_{n}(1-z)^{n}+\delta g (1-z)^{b}\right) \ ,\\
A(z) &= (1-z)^{1/2} \ .
\end{align}
\label{eqn:IRPEDseries}
\end{subequations}
We can consistently set $B\equiv 0$ in our equations of motion and ansatz, and take for the electric component of the Maxwell field and for the dilaton the expansions
\begin{subequations}
\begin{align}
h(z) &= (1-z)\left( \sqrt{\frac{f_{0}}{2}} +\delta h (1-z)^{b} \right) \ , \\
\Phi(z) &= -\frac{\sqrt{3}}{4}\log{(1-z)}+\delta \Phi (1-z)^{b} \ .
\end{align}
\end{subequations}
We have turned on a perturbation $\{\delta f, \delta g, \delta h, \delta \Phi \}$, with $b = \frac{1}{6}\left(-3+\sqrt{57}\right)$, to allow us to flow to different values of the dual field theory parameters in eqns.~\reef{eqn:dimensionlessratios}. To find the perturbation, we treat the equations of motion (EOM) as functions of the $\delta f$, etc., and then consider the linear problem
\begin{equation}
\frac{\partial{\text{EOM}(\delta_{I})}}{\partial{\delta_{J}}}\Big|_{\delta_{J}=0}\delta_{J}=0  \ ,
\label{eqn:perturbations}
\end{equation}
where $I,J=1,\dots,4$ and $\delta_{1}=\delta f, \delta_{2}=\delta g$, etc., and ``EOM'' in eqn.~\reef{eqn:perturbations} refers to each of the eqns.~\reef{eqn:eom}. We require this to have a nontrivial solution for the~$\delta_{I}$ and thus, the matrix of derivatives must be non--invertable. Requiring its determinant to vanish gives us a polynomial condition for~$\beta$ which we can solve, and then using this, we can solve the linear system in eqn.~\reef{eqn:perturbations} for the perturbations.

 Substitution of eqn.~\reef{eqn:IRPEDseries} into the equations of motion in eqt.~\reef{eqn:eom} fixes all coefficients in the expansions except for $f_{0}$, which is chosen so that the UV metric flows to AdS. In fact, this fixes all perturbations except $\delta f$, whose choice takes us to different values of the dimensionless ratios in eqn.~\reef{eqn:dimensionlessratios}.
 
\begin{figure}[h!]
\centering
\includegraphics[width=0.6\textwidth]{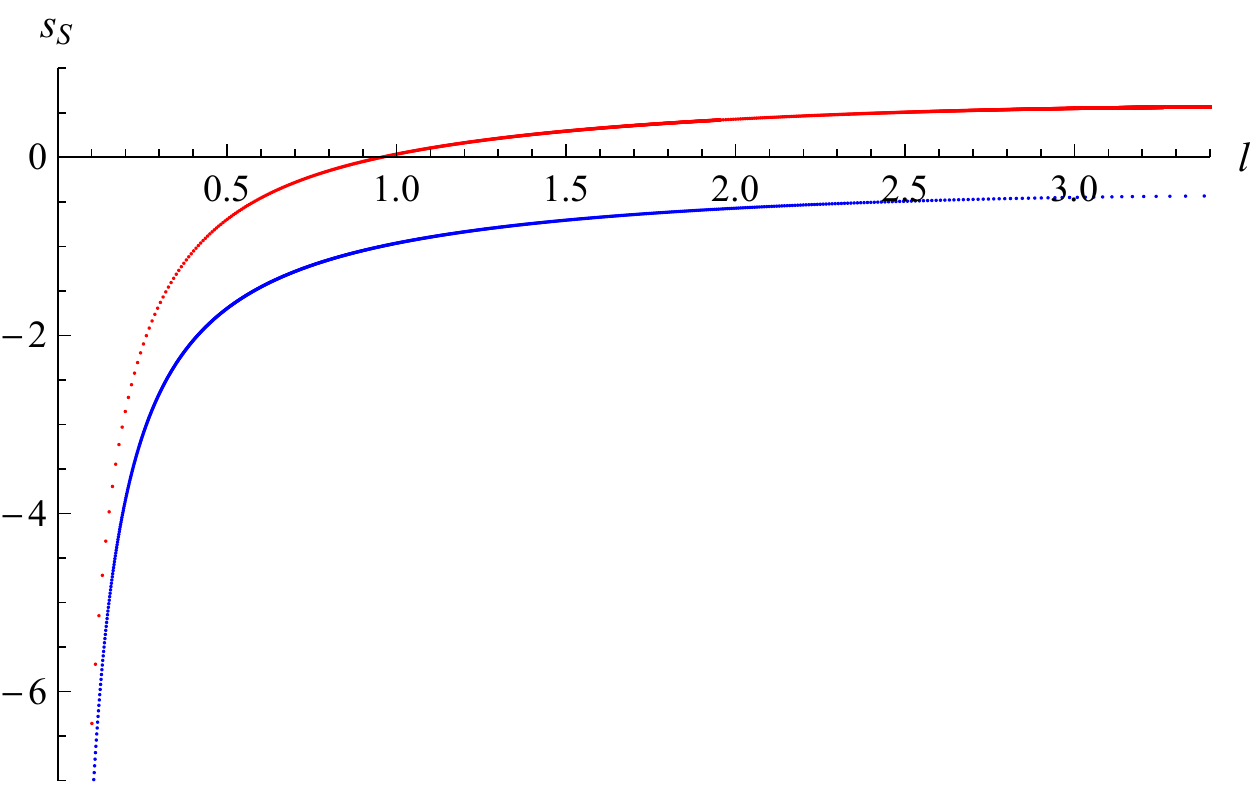}
\caption{The finite part of the entanglement entropy $s_{S}$ as a function of the strip width $\ell$ for the numerical PED (blue/lower), with $\delta f =0$, and the analytical solution (red/upper), eqns. \reef{eqn:Gubsermetric}. We have taken our UV cutoff to be $\epsilon = 10^{-5}$ and our maximal IR value of $z_{T}=0.9999$. See text for explanation of the shift.}
\label{fig:PEDvsGubser}
\end{figure}

 We now compute the finite part of the entanglement entropy for the strip geometry using eqns.~\reef{eqn:EEstripintegral} and \reef{eqn:EEstriplength}. Figure~\ref{fig:PEDvsGubser} shows a plot of $s_{S}$ as a function of the strip width $\ell$. In order to check our numerics, we compared this calculation with that of the analytical solution for the PED in ref.~\cite{Gubser:2009qt}\footnote{In comparing their four--dimensional Lagrangian with ours, the relationships between their dilaton $\alpha$ and ours is $\alpha=\frac{2}{\sqrt{3}}\Phi$. In addition, their metric is in terms of a coordinate $r'$, but we have done a change of coordinates to $z'=L'/r'$ and replaced their $\{t',\vec{x}' \}$ by $L' \{t',\vec{x}'\}$ for comparison. We have also defined a dimensionless charge $\tilde{Q}$ via $\tilde{Q}=Q'/L'$. We have omitted writing expressions for their dilaton and Maxwell field as their particular form is not needed for what we discuss here.}. Their metric, written in the form of eqn.~\reef{eqn:metric}, is
\begin{subequations}
\begin{align}
F(z') &=\left(1+\tilde{Q}z'\right)^{3/2}\left(1-\frac{\tilde{Q}^3z'^{3}}{\left(1+\tilde{Q}z'\right)^3}\right) \ , \\
G(z')&=\left(1+\tilde{Q}z'\right)^{-3/2}\left(1-\frac{\tilde{Q}^3z'^{3}}{\left(1+\tilde{Q}z'\right)^3}\right)^{-1} \, ,\\
A(z')&=\left(1+\tilde{Q}z'\right)^{3/2} \, \ ,
\end{align}
\label{eqn:Gubsermetric}
\end{subequations}
where we have used a prime to denote their coordinates, and in their coordinates $\{t',z',\vec{x}'\}$ the horizon is at $z'=\infty$ and the UV at $z'=0$, and $L'$ is the AdS radius. Lastly, to compare to our solution we did one further rescaling of the coordinates by $\tilde{Q}$ so that eqns. \reef{eqn:Gubsermetric} are only in terms of~$z'$, although in Section \ref{EMDual} we need the explicit $\tilde{Q}$ dependence and so have kept it in for eqns. \reef{eqn:Gubsermetric}.

 The analytic solution corresponds to our $\delta f=0$, for which the dimensionless UV field theory parameters take the values
\begin{equation}
\mu=1.2247 \ ,\qquad \frac{\phi_{1}}{\mu}=0.3535 \ ,\qquad \frac{Q}{\mu^2}=0.8165 \ ,\qquad \frac{\phi_{2}}{\mu^2}=0.1444 \ .
\end{equation}
These values agree with the analytical values to within $10^{-4}$ accuracy. The result of the analytical entanglement entropy computation is also shown in Figure~\ref{fig:PEDvsGubser}, and there seems to be a mismatch. However, this is  due to the difference in coordinates used in the two calculations. The analytic solution's coordinates are related to our coordinates via $z=\frac{z'}{1+z'}$. The consequence is that when regulating the entanglement entropy so that it is finite, we must specify what our chosen regulator is and this is a coordinate dependent statement. When comparing the two results we must take into account that the UV regulators are different because the location of the UV relative to the~IR is rather different in the two cases; indeed, the shift seen in Figure~\ref{fig:PEDvsGubser} for the asymptotic IR value of~$s_{S}$ is 1, which from the coordinate relationship between the two systems is precisely the difference in regulators: $1/\epsilon = 1 + 1/\epsilon'$. 

\begin{figure}[t!]
\centering
\subfigure[]{\includegraphics[width = 3in]{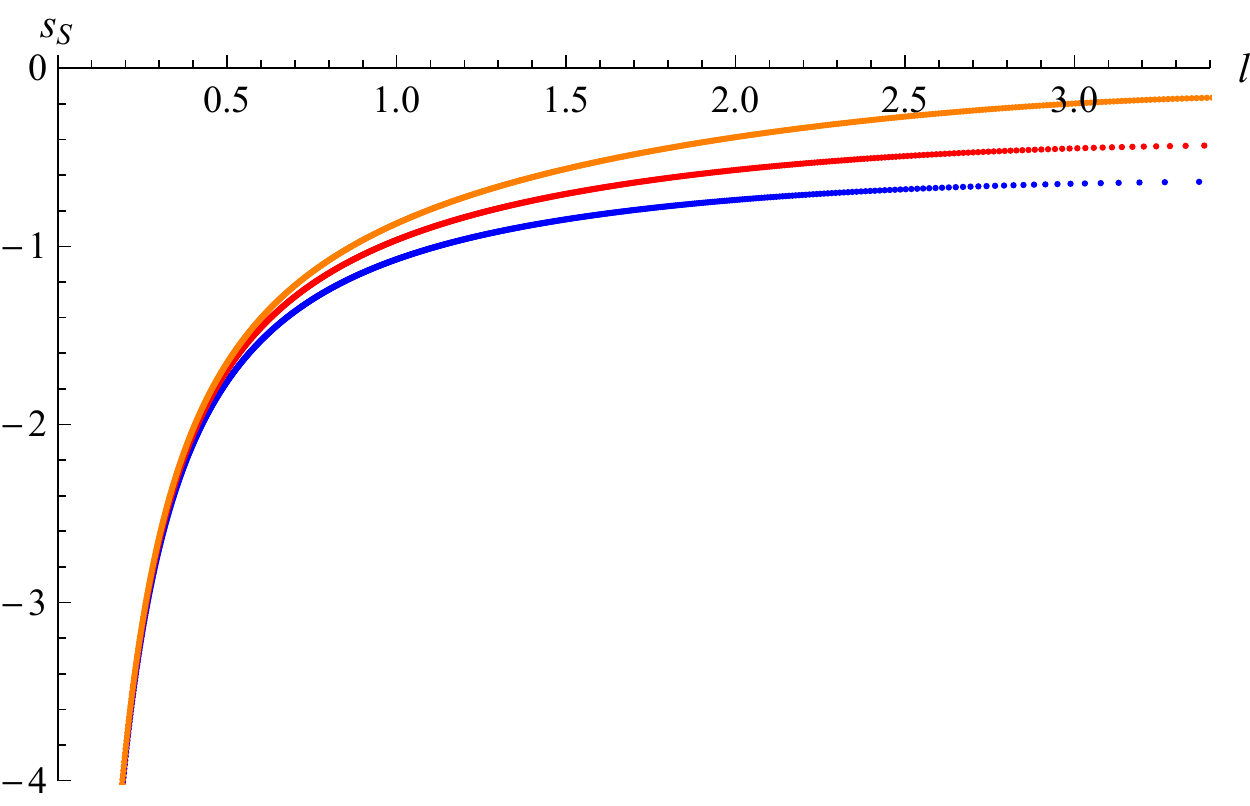}\label{fig:EEmultiPED}} 
\subfigure[]{\includegraphics[width = 3in]{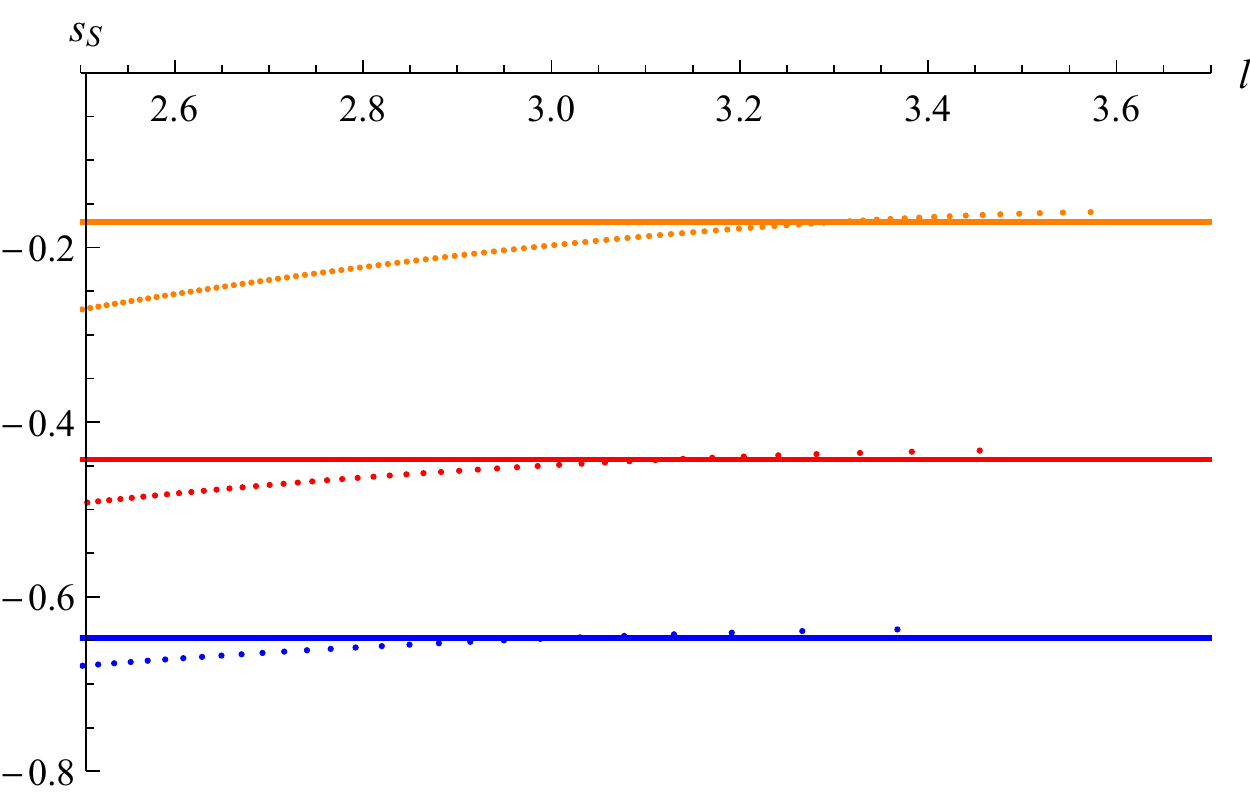}\label{fig:EEmultiPEDzoomed}}
\caption{(a) The finite strip entanglement entropy $s_{S}$ as a function of $\ell$ for various values of $\delta f$. Blue/lower is $\delta f = -10$ ($Q/\mu^2=1.0209$), red/middle is $\delta f = 0$ ($Q/\mu^2=0.8165$), and orange/upper is $\delta f = 10$ ($Q/\mu^2=0.5863$). (b) A zoomed in version of (a) where the horizontal lines are the areas of the in--falling solution for each.}
\end{figure}

 It is usually expected that the large $\ell$ limit of the entanglement entropy should approach the thermal entropy of the system. However, in this case, our PED solutions have zero thermal entropy and we see that both the numerical and the analytic solution asymptote to two different values as $\ell$ increases. In fact, we see in Figure~\ref{fig:EEmultiPED} plots of the strip entanglement entropy for various values of $\delta f$ each of which asymptotes to a different large $\ell$ value. We also find that as $\delta f$ is increased, the value of the dual field theory chemical potential $\mu$ also increases, while the values of $\phi_{1}/\mu$ and $Q/\mu^2$ decrease.
 
\begin{figure}[h!]
\centering
\includegraphics[width=0.6\textwidth]{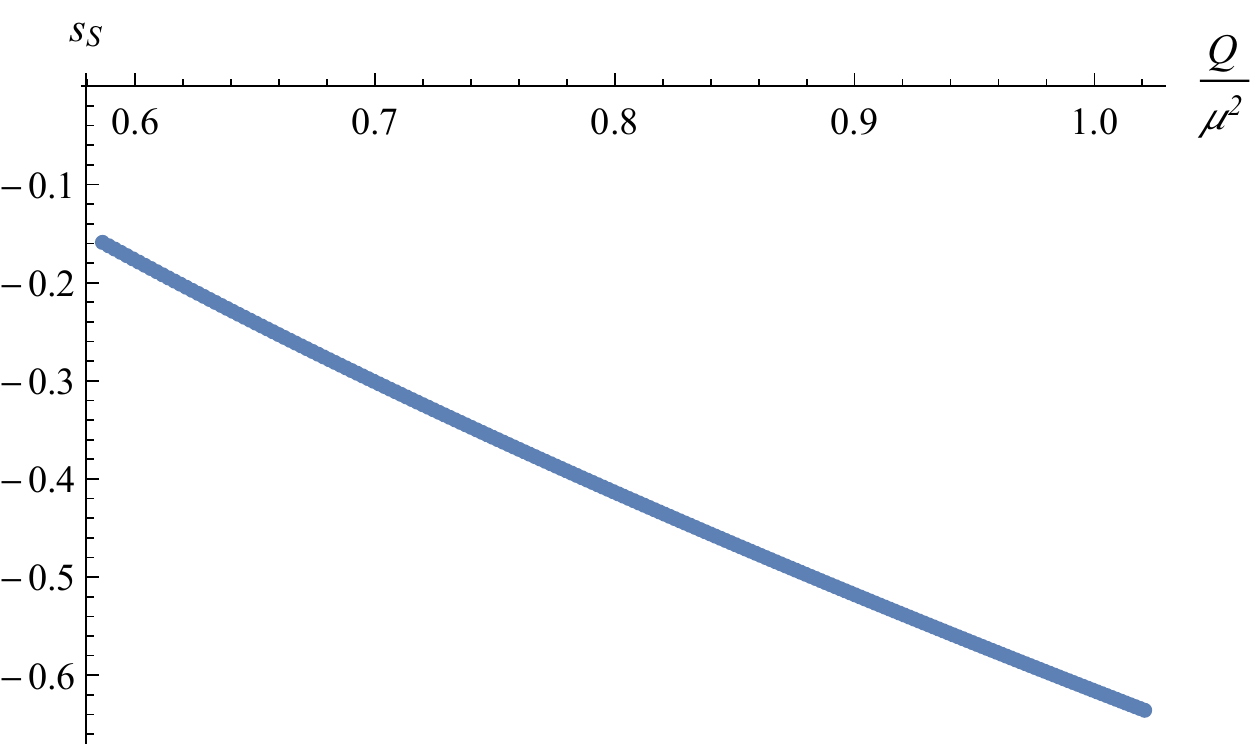}
\caption{The large $\ell$ value of $s_{S}$ as a function of $Q/\mu^2$ for $-10\leq \delta f \leq 10$.}
\label{fig:ShiftvsQ}
\end{figure}

  Interestingly, if we consider the ``in--falling'' minimal surface, namely the one that just hangs straight down into the black hole, and compute its finite area via
\begin{equation}
\frac{\text{Area}(\gamma_{\text{if}})}{2\mathscr{L}}=\int_{\epsilon}^{1}{dz\sqrt{a(z)g(z)}}-\frac{1}{\epsilon} \ ,
\end{equation}
(where $\gamma_{\text{if}}$ denotes the in--falling surface) we find that this value agrees to within reasonable numerical precision with the large $\ell$ value of $s_{S}$. For example, when $\delta f = 0$, we find that the large~$\ell$ value of $s_{S}$ is given by $-0.4325$ and the finite area of the in--falling piece is given by $-0.4427$. This suggests that these ``side pieces'' contribute to the entanglement entropy. Since our solutions have zero temperature and zero thermal entropy, as well as a zero area horizon, this may be the reason we are able to see their contribution, whereas in the case of a non--zero thermal entropy, these pieces are negligible compared to the thermal entropy. Figure~\ref{fig:EEmultiPEDzoomed} shows the same plots as Figure~\ref{fig:EEmultiPED} but zoomed in to show a comparison between the large $\ell$ behavior of the entanglement entropy and the in--falling piece. In Figure~\ref{fig:ShiftvsQ}, we show how the large $\ell$ value of $s_{S}$ changes as the dimensionless ratio $Q/\mu^2$ is varied.

\begin{figure}[h!]
\centering
\includegraphics[width=0.6\textwidth]{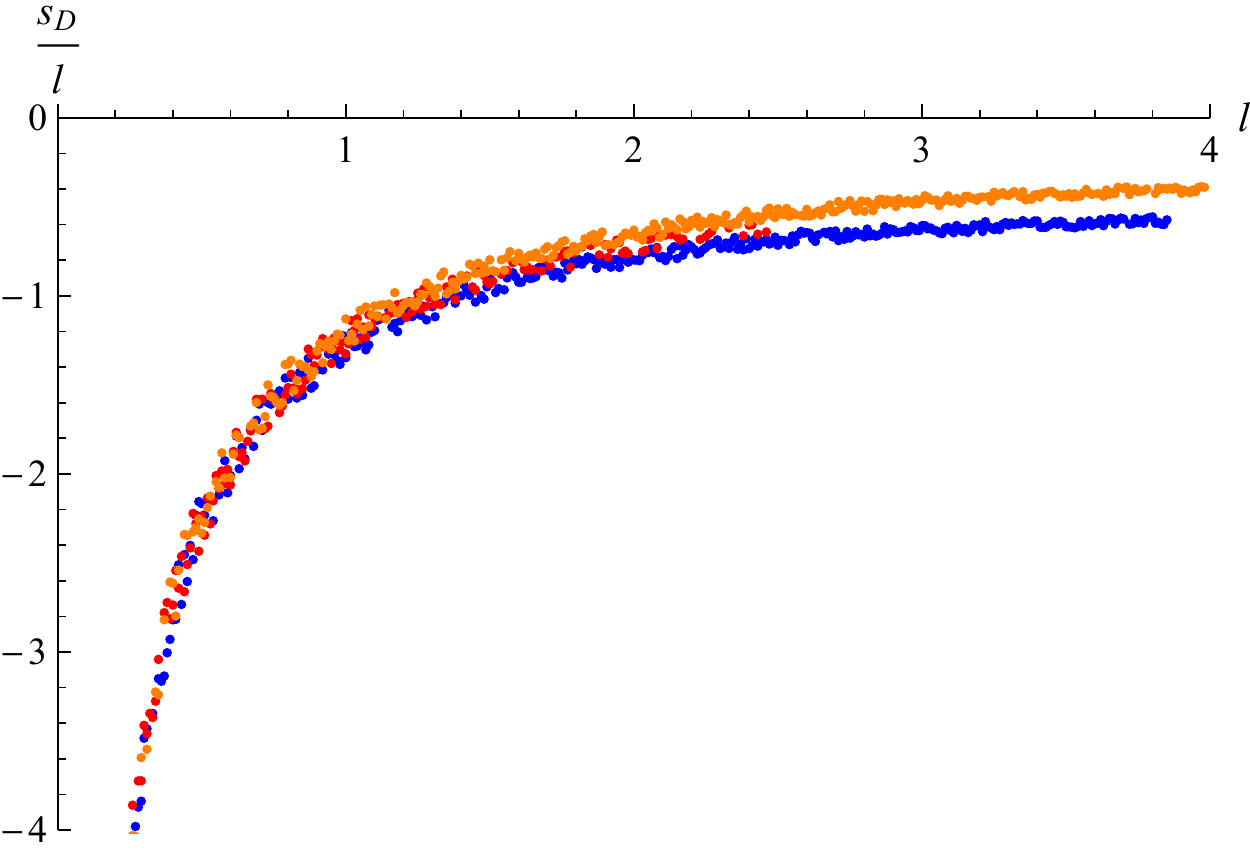}
\caption{The finite part of the entanglement entropy $s_{D}$ as a function of the disk radius $\ell$ for three values of $\delta f$. Blue/lower is $\delta f = -4$ ($Q/\mu^2=0.9002$), red/middle is $\delta f = 0$ ($Q/\mu^2=0.8165$), and orange/upper is $\delta f = 4$ ($Q/\mu^2=0.7289$). Recall that finding $s_{D}$ requires solving a second order differential equation as well as numerical integration, so the numerical accuracy is not as controlled as in the strip case, hence the numerical variance in the entanglement entropy curves shown.}
\label{fig:EEdiskPED}
\end{figure}

 We can also compute the entanglement entropy for the disk geometry $s_{D}$ defined in eqn.~\reef{eqn:EEdisk} and found by computing the integral in \reef{eqn:EEdiskpullback} using the minimal surface $z(r)$ found by extremizing the area functional of \reef{eqn:EEdiskpullback}. For numerical reasons, we find it more convenient to work with a new radial coordinate defined by \cite{Albash:2011nq}
\begin{equation}
\zeta(\ell) = \frac{1}{\ell}\sqrt{\ell^2+\epsilon^2-r^2} \ ,
\label{eqn:radialcoord}
\end{equation}
where $\ell$ is the radius of the disk. In terms of this radial coordinate the pure $\text{AdS}_{4}$ minimal surface would just be given by $z(\zeta)=\ell\zeta$. Using this, we are able to find $s_{D}$ for various values of $\delta f$. In Figure~\ref{fig:EEdiskPED}, we plot $s_{D}/\ell$ vs. $\ell$ for three different values of $\delta f$. We see\footnote{\label{fn:numvariance}As mentioned in Section \ref{EEReview} finding $s_{D}$ requires solving a order differential equation as well as numerical integration. Since our backgrounds are numerical and we also have to solve the second order differential equation numerically, we have some numerical variance, as seen in the disk plots.} that entanglement entropies seem to be asymptoting to different large $\ell$ values as in the case with the strip.

\section{Entanglement Entropy of the Purely Magnetic Dilatonic Black Hole} \label{PMD}
In this section, we will consider dilatonic black holes that have no electric charge but do have a magnetic charge. We can use as our ansatz the same metric \reef{eqn:metric} and Maxwell fields \reef{eqn:maxwell} except we set $h(z)\equiv 0$. The IR expansion for our metric fields and dilaton then becomes
\begin{subequations}
\begin{align}
f(z) &= \frac{(1-z)^2}{z^2}\sum_{n=0}^{\infty}{f_{n}(1-z)^{4n/3}} \ ,\\
g(z) &= (1-z)^{-4/3}\sum_{n=0}^{\infty}{g_{n}(1-z)^{4n/3}} \ ,\\
a(z) &= (1-z)^{2/3}\sum_{n=0}^{\infty}{a_{n}(1-z)^{4n/3}} \ ,\\
\Phi(z) &=\frac{\sqrt{3}}{3}\log{(1-z)}+\sum_{n=0}^{\infty}{\Phi_{n}(1-z)^{4n/3}} \ .
\end{align}
\end{subequations}
We find that the entire function $f(z)$ is free in the IR, along with the values of $B$ and $\Phi_{0}$. For simplicity, we fix $f(z) = (1-z)^2/z^2$ and choose $B$, tuning the value of $\Phi_{0}$ so that the other metric functions give the correct $\text{AdS}_{4}$ behavior. In this case, we do not find the need to introduce a perturbation to allow us to flow to different values of the UV parameter that characterize the theory. Since here we do not have any electric component to our solution, we will take the dimensionless ratio to be $B/\phi_{1}^2$.

\begin{figure}[h!]
\centering
\subfigure[]{\includegraphics[width=3in]{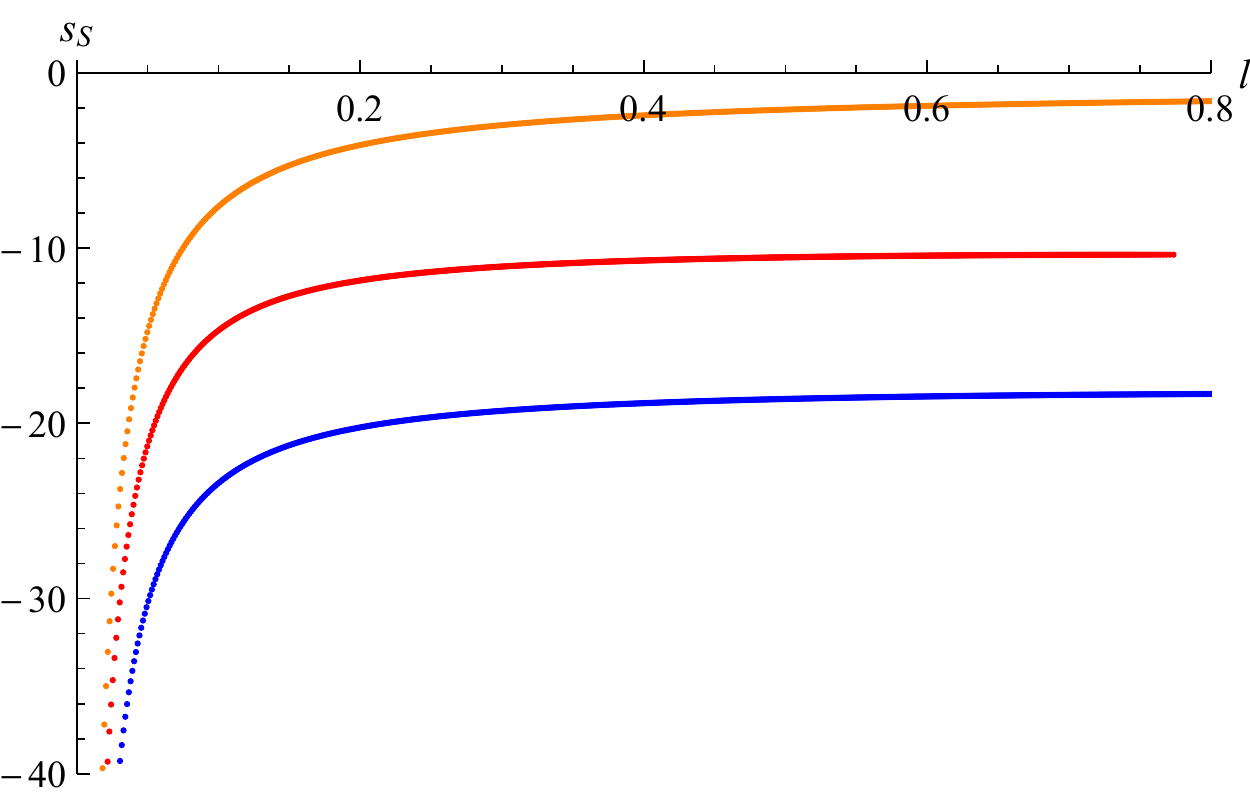} \label{fig:EEstripPMD}}
\subfigure[]{\includegraphics[width=3in]{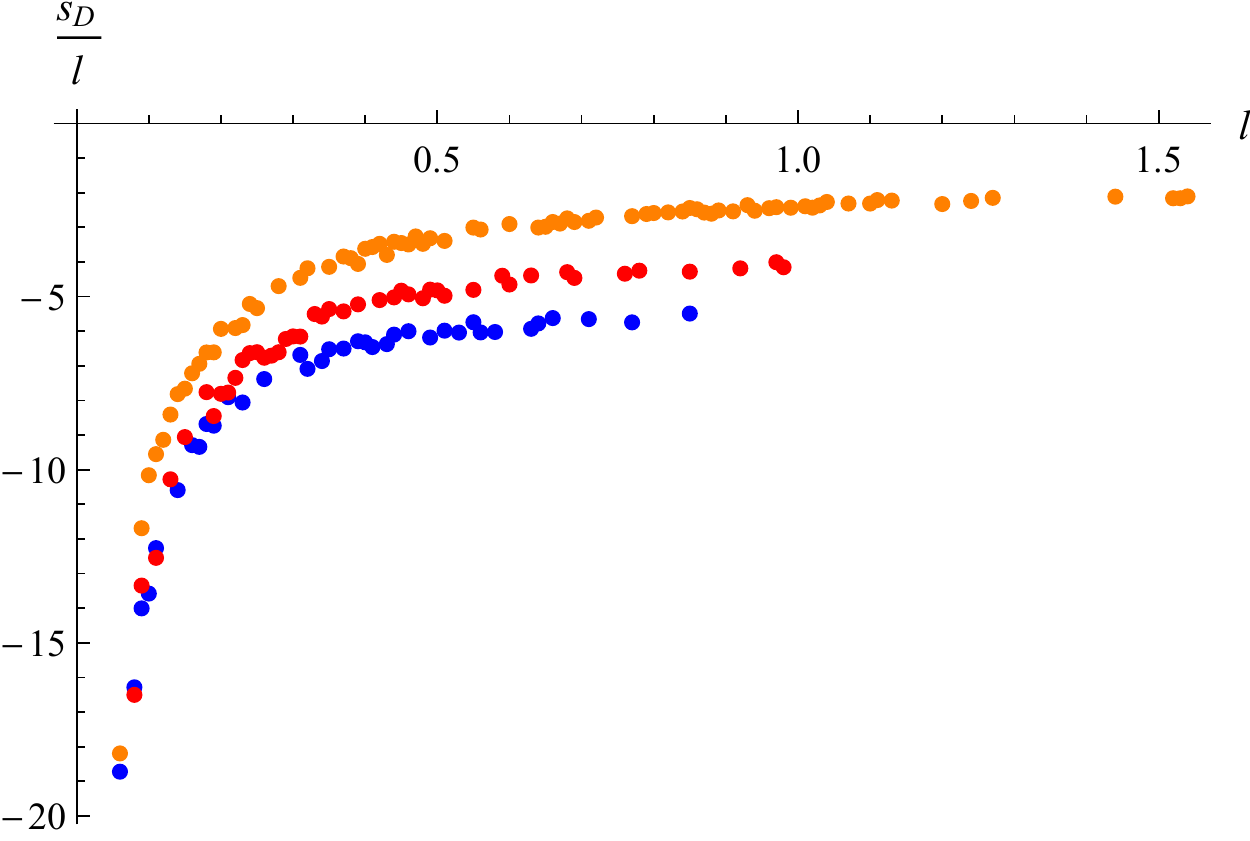}\label{fig:EEdiskPMD}}
\caption{(a) The finite part of the entanglement entropy $s_{S}$ as a function of the strip width $\ell$ for the PMD with $B/\phi_{1}^2=0.4183$ (blue/lower), $B/\phi_{1}^2=0.6710$ (red/middle), and $B/\phi_{1}^2=1.6383$ (orange/upper). (b) The finite part of $s_{D}/\ell$ as a function of $\ell$ for the same value of the UV parameters with the same colors/locations. See footnote \ref{fn:numvariance} for an explanation of the numerical variance in (b).}
\end{figure}

 Figure~\ref{fig:EEstripPMD} shows the finite part of the entanglement entropy for the strip geometry for three solutions with different values of $B/\phi_{1}^2$. We note that the behavior seems similar to that of the PED, suggesting that the form of the entanglement entropy may be invariant under electromagnetic duality. We explore this next.
 
 For the disk geometry, the finite part of the entanglement entropy can be found in Figure~\ref{fig:EEdiskPMD} for the same values of the UV parameter. Again, the behavior is similar as to that of the PED disk entanglement entropy.

\section{Electromagnetic Duality} \label{EMDual}
For both the PED and PMD solutions, the Maxwell equation of motion from the action in eqn.~\reef{eqn:EMDaction} is 
\begin{equation}
\partial_{\nu}\left(\sqrt{-G}Z\left(\Phi\right)F^{\mu\nu}\right)=0 \ .
\label{eqn:maxwelleom}
\end{equation}
In vacuum, there is the well--known electromagnetic duality where the equation of motion, eqn.~\reef{eqn:maxwelleom}, is invariant under $F_{\mu\nu}\rightarrow \star F$, where $\star F$ is the Hodge dual given by
\begin{equation}
\left(\star F\right)_{\mu\nu}= \frac{1}{2}\sqrt{-G}\varepsilon_{\mu\nu\rho\sigma}G^{\rho\lambda}G^{\sigma\omega}F_{\lambda\omega}.
\label{eqn:maxwelldual}
\end{equation}
Here, $\varepsilon_{\mu\nu\rho\sigma}$ is the completely antisymmetric tensor with $\varepsilon_{0123}=+1$. In the case when there is a scalar in the theory that couples to the Maxwell field strength as in the action in eqn.~\reef{eqn:EMDaction} as $Z(\Phi)F^2$, then the electromagnetic duality also requires that $\Phi(z)\rightarrow - \Phi(z)$. We now consider the electromagnetic duality in the boundary theory, applied to the PED and PMD solutions and their strip entanglement entropy.

Evaluating eqn.~\reef{eqn:maxwelldual} for the PED solution with metric written as in eqn.~\reef{eqn:IRPEDseries} we have that
\begin{equation}
\left(\star F\right)_{xy}=-\frac{A(z)}{\sqrt{F(z)G(z)}}h'(z) \ .
\end{equation}
Recall we require the UV behavior of $h(z)=\mu - Qz$ and $\{F,G,A\}\rightarrow\{1,1,1\}$. Thus, the UV ``magnetic dual'' solution to the PED solution with given $Q$ value is
\begin{equation}
\left(\star F\right)_{xy}=Q \ .
\end{equation}
Because we have scaled out the dependence of the solution on the horizon, $z_{H}$, we must always consider dimensionless ratios. Thus in comparing the PED and PMD solutions under electromagnetic duality, we fix the value of $Q/\phi_{1}^{2}$ in the PED and then find the corresponding PMD solution that has $B/\phi_{1}^{2}=Q/\phi_{1}^{2}$. We also check that the UV behavior of the dilaton has opposite sign between the two theories.

\begin{figure}[h!]
\centering
\subfigure[]{\includegraphics[width=3in]{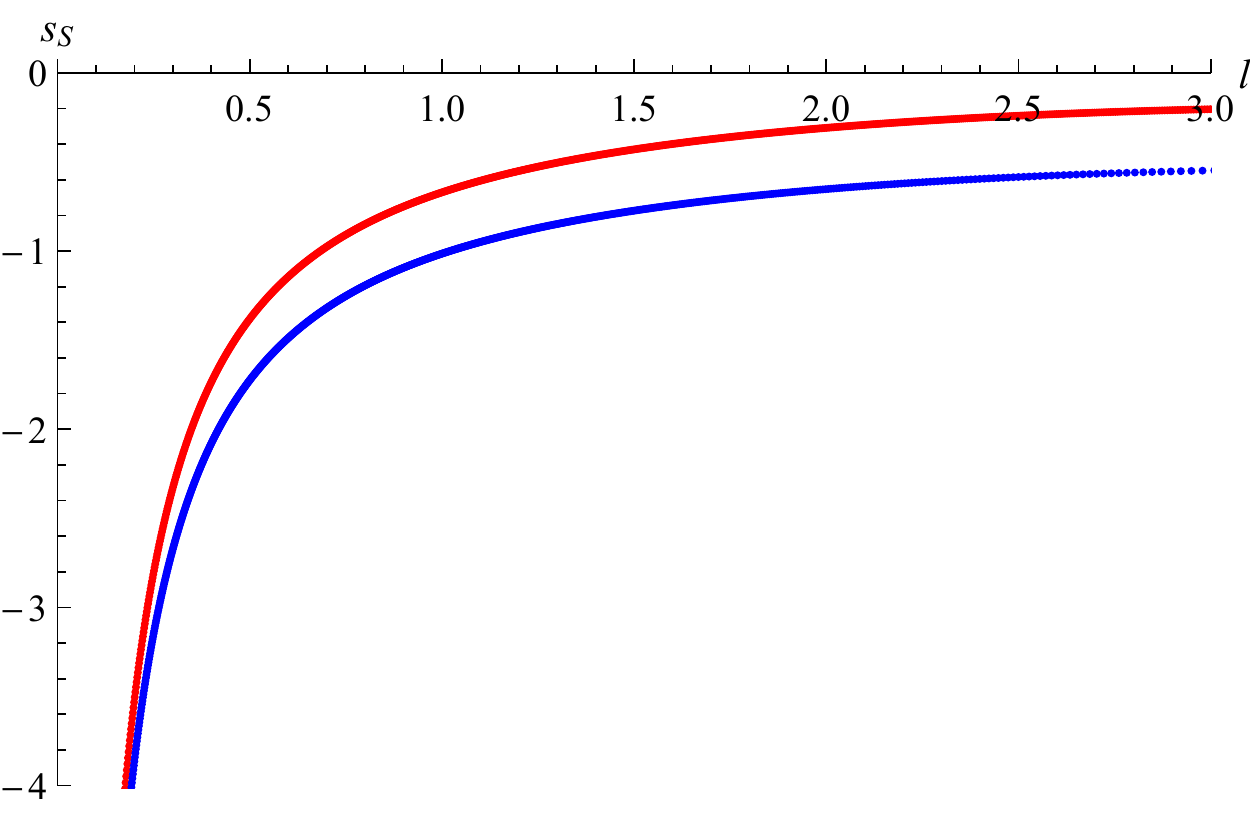} \label{fig:EMdual}}
\subfigure[]{\includegraphics[width=3in]{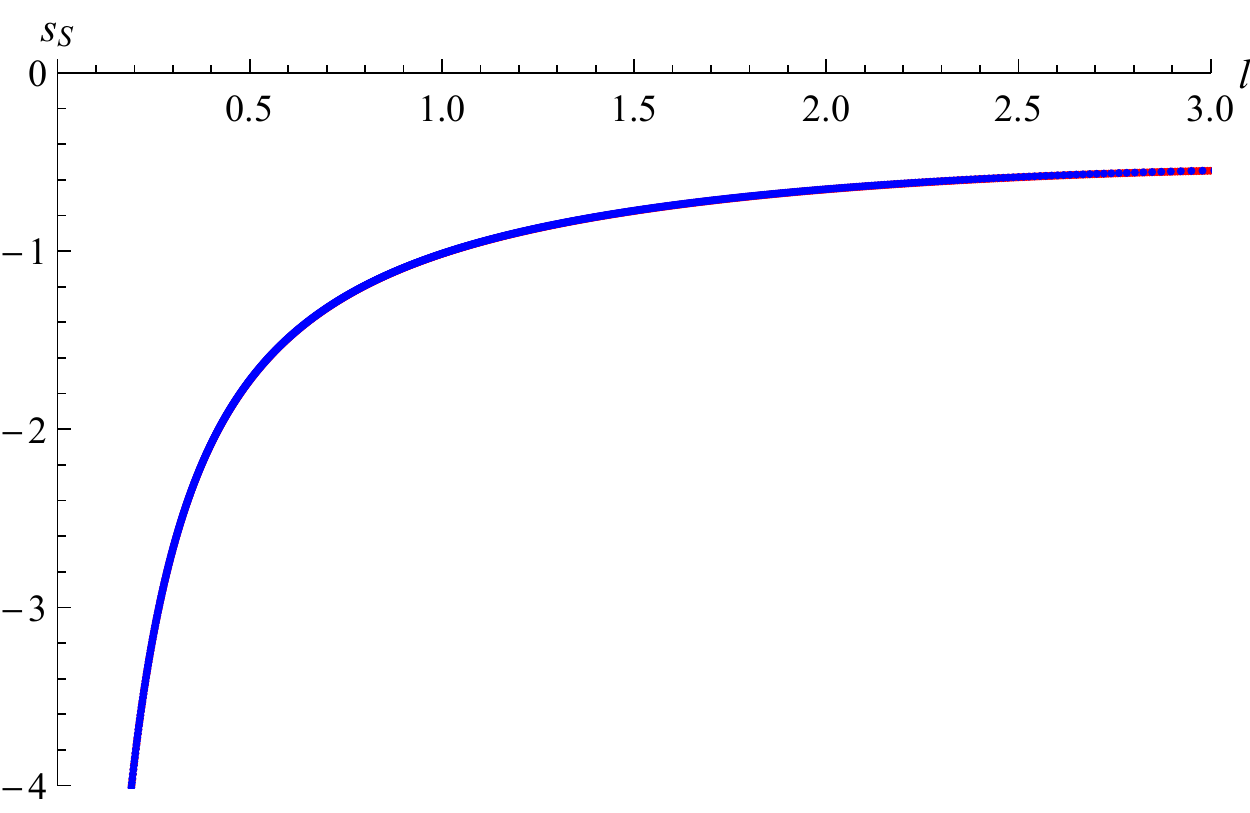}\label{fig:EMdualshifted}}
\caption{(a) The finite part of the strip entanglement entropy $s_{S}(\ell)$ for the PMD (red/upper) and the PED (blue/lower), EM dual solutions with ratios $B/\phi_1^2=Q/\phi_1^2=3.4068$. The PMD is shifted up by $0.3440$. (b) The same entanglement entropies as in (a) but with the PMD shifted down; see text for explanation. The agreement is nearly perfect.}
\end{figure}

 It turns out that all of our PED solutions have approximately the same value of $Q\approx1.224$, although the dimensionless ratios change. We pick the value $Q/\phi_1^2=3.40684$, which is for a $\delta f$ value of $\delta f = -4.7218$. We then find the corresponding PMD solution with $B/\phi_1^2=3.40688$. We note that for the PED solution, $\phi_1 = 0.5995$ and for the PMD $\phi_1 = -0.5994$. We then compute the finite part of the strip entanglement entropy for both and obtain the results in Figure~\ref{fig:EMdual}. 
 
 We find that the two entanglement entropies are shifted relative to one another, with the PMD solution being $0.3440$ higher than the PED. We observed a similar shift between the analytic and numerical backgrounds for the PED solution $\delta f=0$, which was due to different coordinates giving rise to different UV cutoffs. For the PED and PMD solutions, we have used the same coordinates with the horizon at $z=1$; however, there is a subtlety. We rescaled the radial coordinate~$z$ by the position of the horizon, $z_{H}$, so that we could use the range $0\leq z \leq 1$ for our numerics. However, in the definition of the entanglement entropy in eqn.~\reef{eqn:EEstripintegral}, this rescaling will change the UV cutoff $\epsilon$ by $\epsilon/z_{H}$. Thus, if we are to restore the position of the horizon, we would have
\begin{subequations}
\begin{align}
\epsilon_{\text{PED}} &= \epsilon z_{Q} \ ,\\
\epsilon_{\text{PMD}} &= \epsilon z_{B} \ ,
\end{align}
\label{eqn:UVcutoffs}
\end{subequations}
where $\epsilon = 10^{-5}$ is the numerical UV cutoff we have used in both backgrounds.  If our choices of the PED horizon $z_{Q}$ and the PMD horizon $z_{B}$ are not the same, it would lead to slightly different~UV cutoffs between the two theories, and hence a shift in the entanglement entropy. 

To see this, recall from the analytic vs. numerical shift we observed previously, we know that the shift, $\eta$, is related to the UV cutoffs via
\begin{equation}
\frac{1}{\epsilon_{\text{PMD}}}=\eta + \frac{1}{\epsilon_{\text{PED}}} \ .
\end{equation}
Using \reef{eqn:UVcutoffs}, this means that the PMD horizon $z_{B}$ in terms of the shift $\eta$ and the PED horizon $z_{Q}$ is given by
\begin{equation}
z_{B}=\frac{z_{Q}}{1+\epsilon \eta z_{Q}}  \ .
\end{equation}
Thus, if the shift is non--zero, then we have that $z_{B}\neq z_{Q}$. We check this by shifting the PMD entanglement entropy down by $s_{S}(\ell)-0.3440$ for all values of $\ell$ and find near perfect agreement with the PED entanglement entropy -- see Figure~\ref{fig:EMdualshifted}. 

\begin{figure}[h!]
\centering
\subfigure[]{\includegraphics[width=3in]{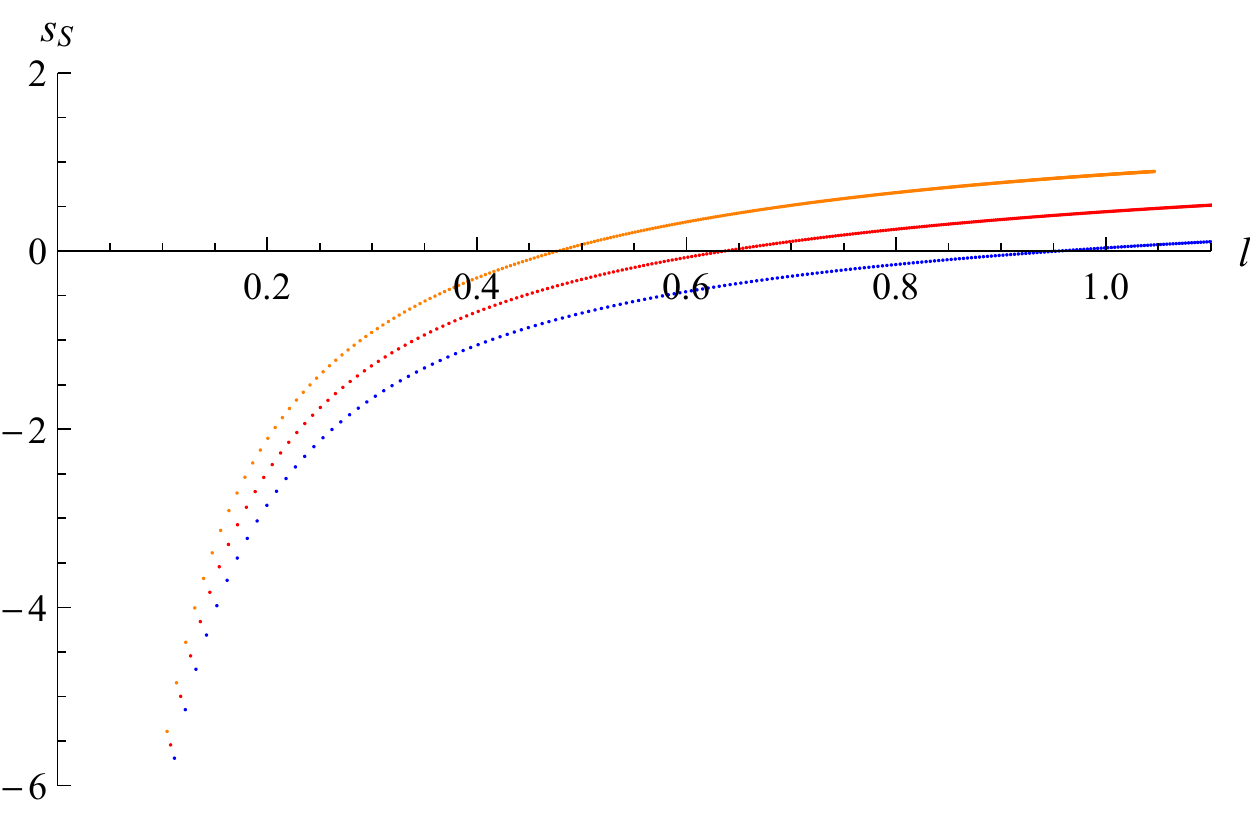} \label{fig:GubserQs}}
\subfigure[]{\includegraphics[width=3in]{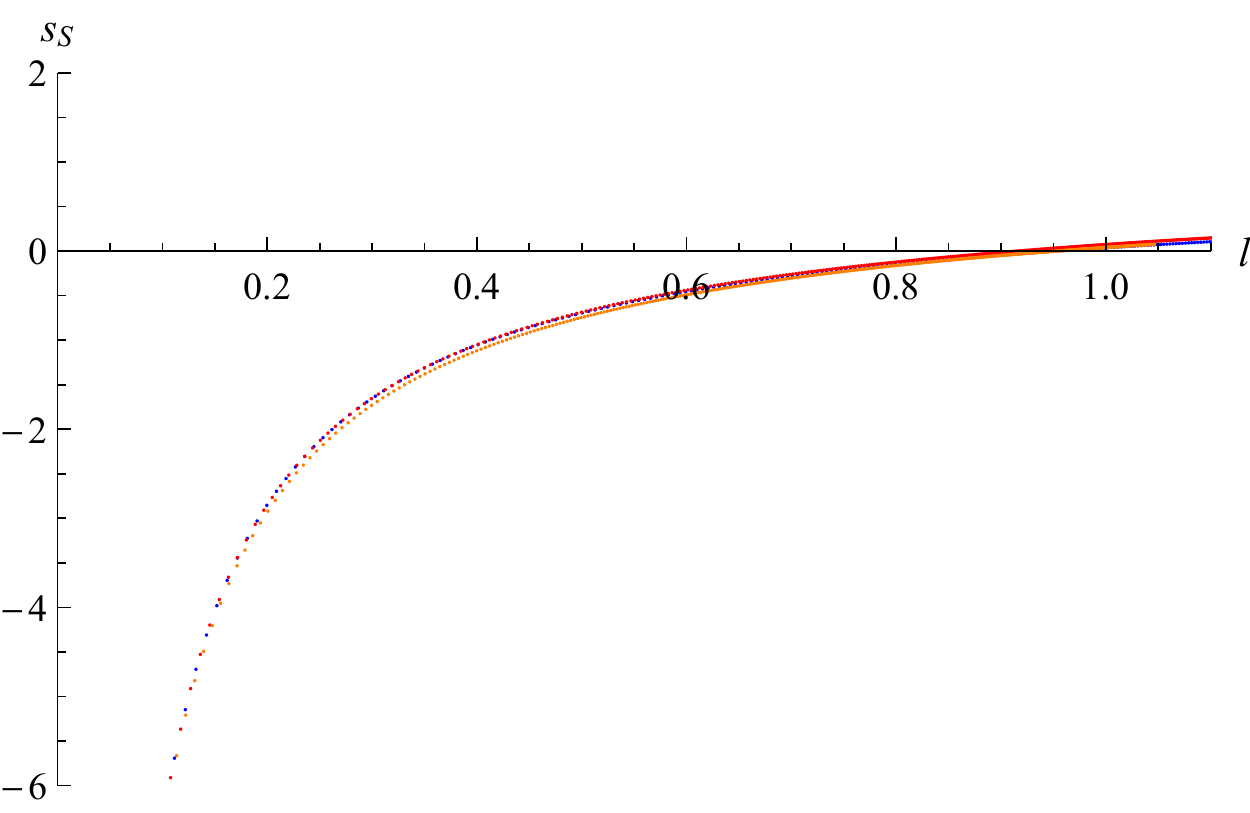}\label{fig:GubserQsShifted}}
\caption{(a) The finite part of the strip entanglement entropy $s_{S}(\ell)$ for the analytic PED solution. The three curves are for $\tilde{Q}=1$ (blue/lower), $\tilde{Q}=1.5$ (red/middle), and $\tilde{Q}=2$ (orange/upper). (b) The same entanglement entropies as in (a) after being shifted down; see text for explanation.}
\end{figure}

 As a further check, we revisit the analytical solution \cite{Gubser:2009qt} for the PED, given in our notation by eqns. \reef{eqn:Gubsermetric}. We can now do a rescaling of the analytic solution's coordinates, letting $z'=\Hat{z}/\tilde{Q}$, which mimics the rescaling of our coordinates by the horizon radius. This implies that the UV cutoff $\epsilon'$ for the analytic solution is then rescaled to become $\Hat{\epsilon}/\tilde{Q}$ and we find a relationship similar to eqn.~\reef{eqn:UVcutoffs}: $\Hat{\epsilon}=\epsilon\tilde{Q}$. Following the above arguments for our case with the different horizons, we find that if $\tilde{Q}$ is varied then there will also be a shift $\eta'$ in the analytic entanglement entropy, controlled by $\tilde{Q}$, via 
\begin{equation}
\tilde{Q}=\frac{1}{1+\epsilon'\eta'}.
\end{equation}
This shows that changing $\tilde{Q}$ should have the same effect as changing the horizon positions in the PED and PMD solutions, which leads to a shift in the entanglement entropy. We show in Figure~\ref{fig:GubserQs} the analytic entanglement entropy $s_{S}(\ell)$ for values of $\tilde{Q}=1, 1.5, 2$ and indeed there is a shift. In Figure~\ref{fig:GubserQsShifted} we subtract this shift from the entanglement entropies when $\tilde{Q}>1$ and we find there is good agreement, which confirms that changing $\tilde{Q}$--and hence changing the horizon position for the PED and PMD solutions--leads to an overall shift.

 From the point of view of the dual field theory, the electromagnetic duality should act by exchanging electric and magnetic charge carriers \cite{Witten:2003ya,Hartnoll:2007ih}, and so our results suggest that the entanglement entropy is invariant under this exchange, at least for the field theories dual to the PED and PMD backgrounds\footnote{In ref.~\cite{Kundu:2012jn}, they also find that under electromagnetic duality transformation, the logarithmic violation of the area law that they see is preserved, suggesting a Fermi surface even though the charge density vanishes.}.

\section{Entanglement Entropy of the Dilaton--Dyon Black Hole} \label{DD}

\begin{figure}[h!]
\centering
\subfigure[]{\includegraphics[width=3in]{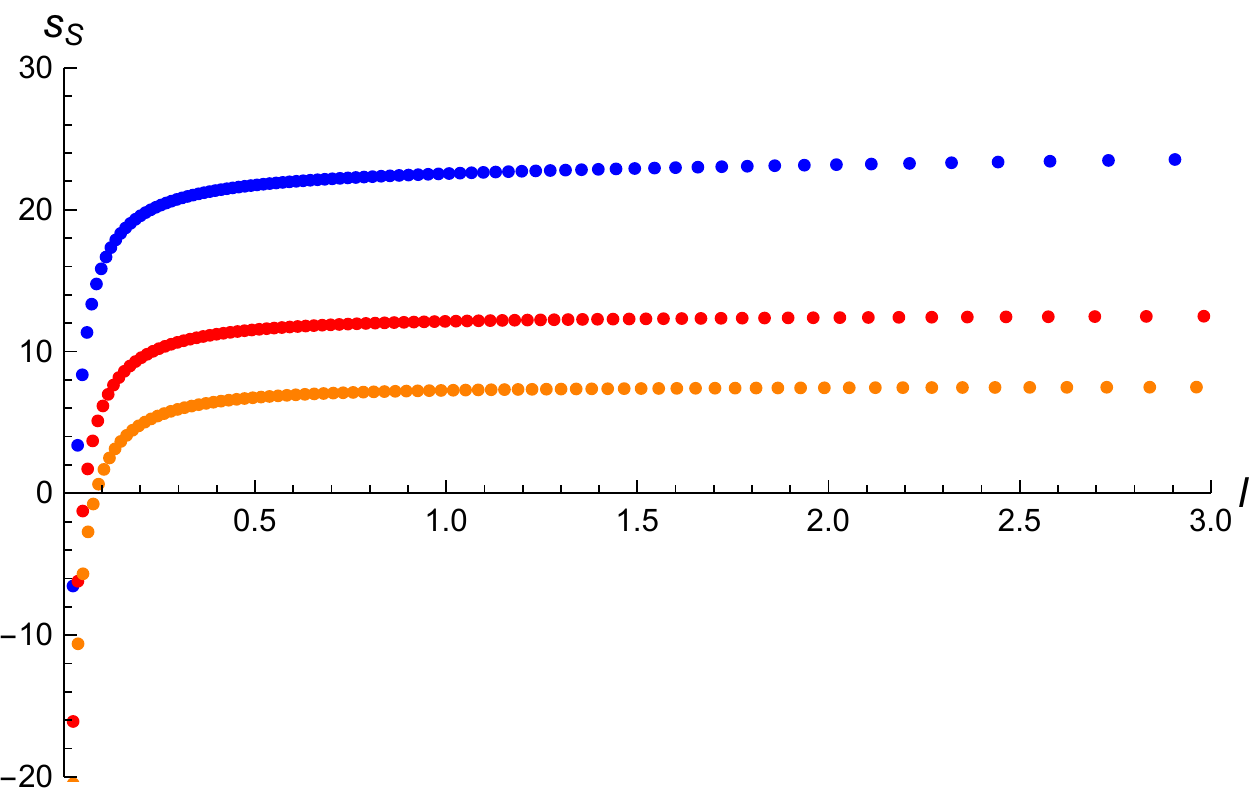}\label{fig:EEstripDD}}
\subfigure[]{\includegraphics[width=3in]{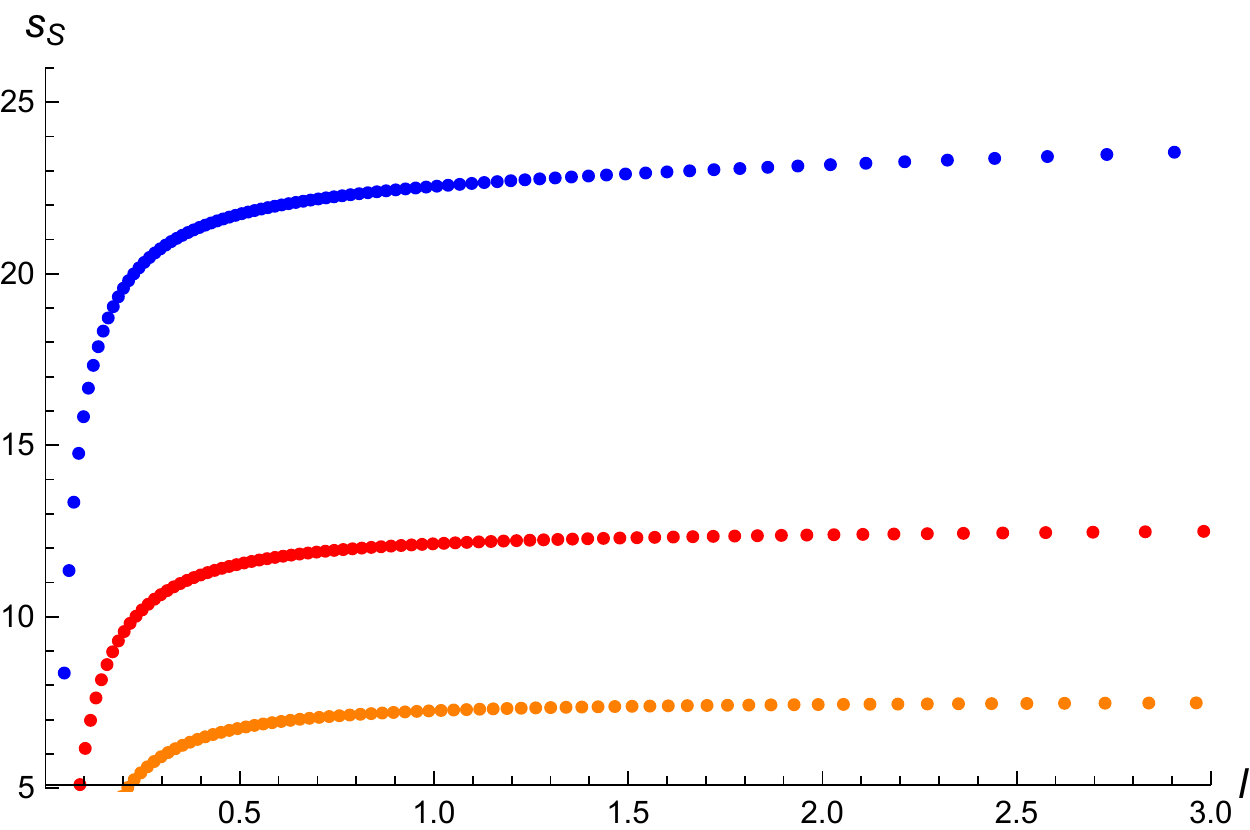}\label{fig:EEstripDDzoomed}}
\caption{(a) The finite part of the entanglement entropy $s_{S}$ as a function of the strip width $\ell$ for three values of the dimensionless ratio $B/\mu^2$. Blue/upper curve is $B/\mu^2=1.040$, red/middle curve is $B/\mu^2=6.779$ and orange/lower curve is $B/\mu^2=27.866$. (b) The same plot as (a) but zoomed in to see the large $\ell$ behavior of $s_{S}$.}
\end{figure}

\begin{figure}[h!]
\centering
\subfigure[]{\includegraphics[width=3in]{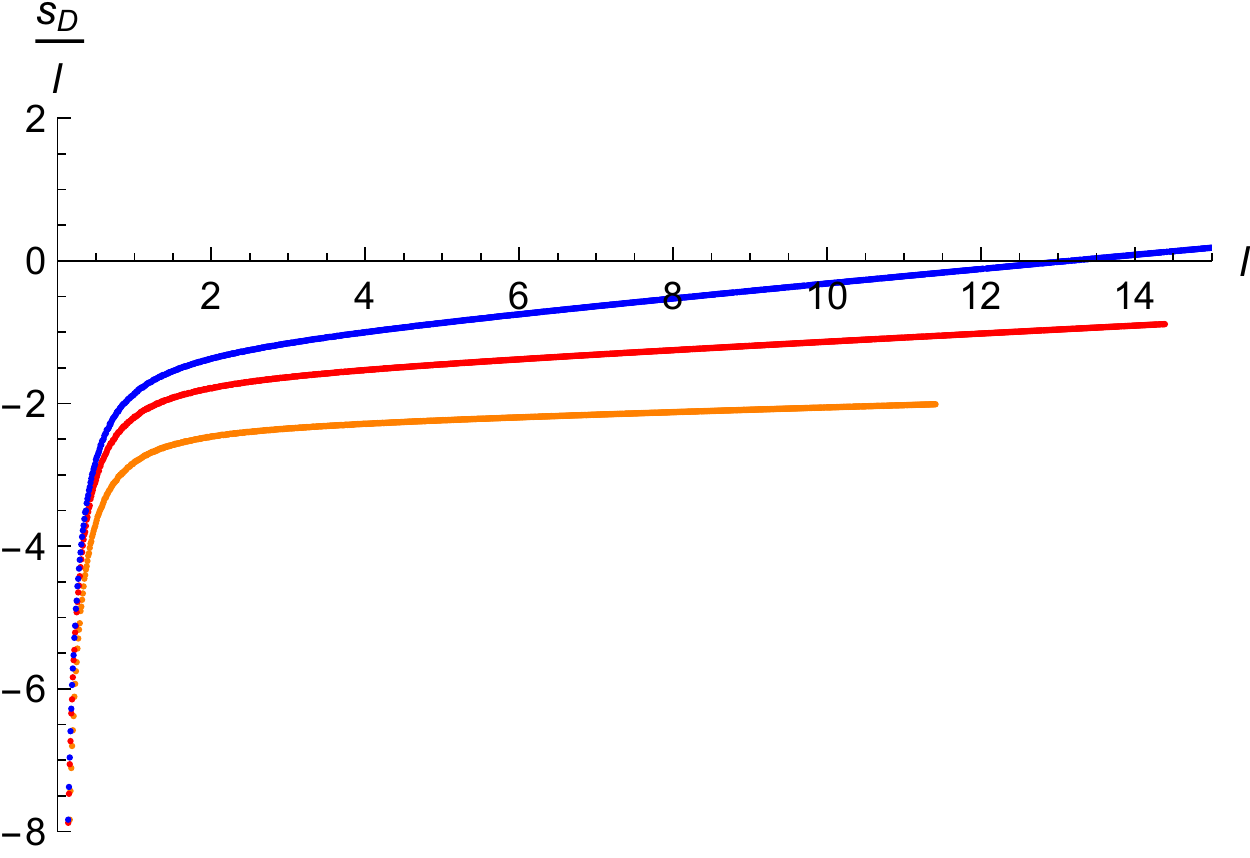}\label{fig:EEdiskDDoverell}}
\subfigure[]{\includegraphics[width=3in]{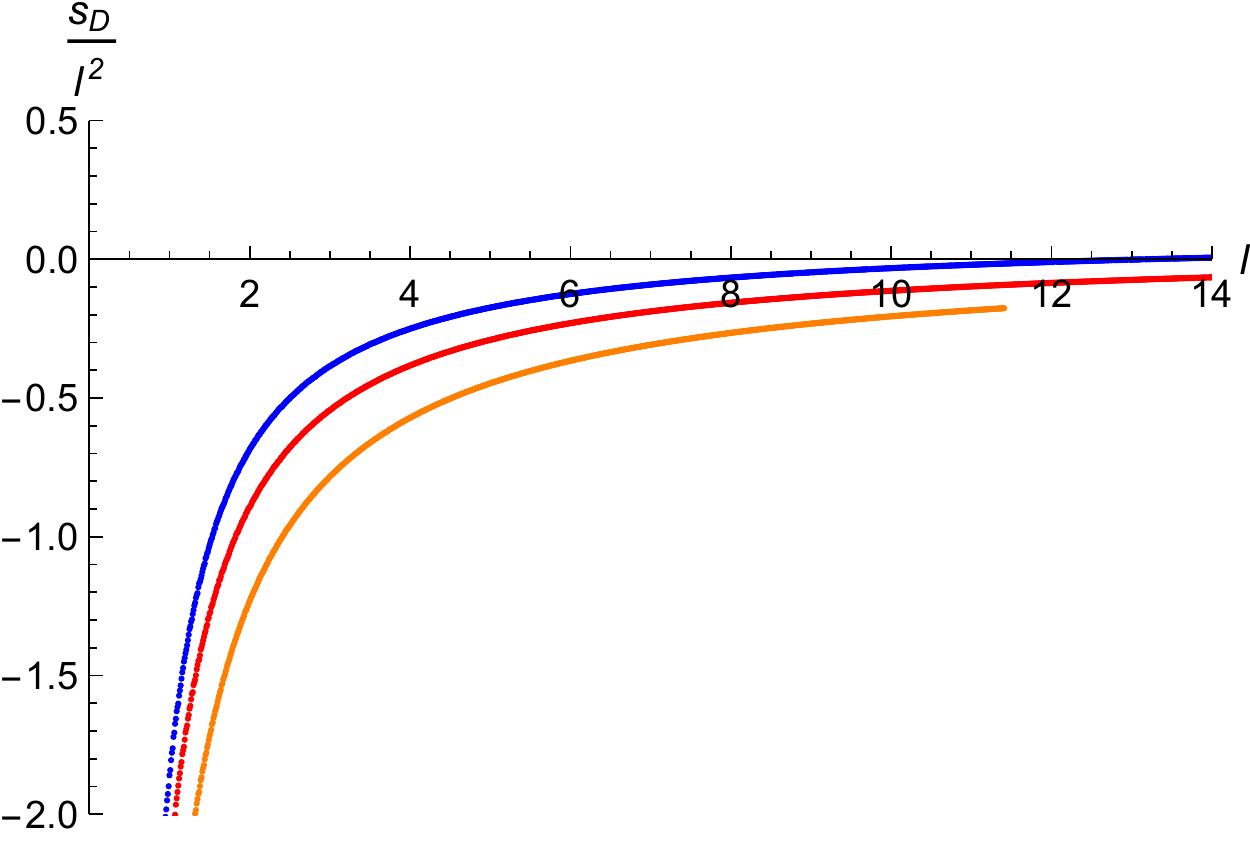}}
\caption{(a) The finite part of the entanglement entropy $s_{D}/\ell$ as a function of the disk radius $\ell$ for four values of the dimensionless ratio $B/\mu^2$. Blue/upper curve is $B/\mu^2=3.249$, red/middle curve is $B/\mu^2=6.779$, and orange/lower curve is $B/\mu^2=27.866$. (b) The same data as in (a) but plotting $s_{D}/\ell^2$. See footnote \ref{fn:DDdisk} for an explanation as to why the results have less numerical variance than those for the PED and PMD.}
\label{fig:EEdiskDD}
\end{figure}

Now we turn to the dilaton--dyon black hole. The IR series expansion for the metric in eqn.~\reef{eqn:metric} Maxwell field \reef{eqn:maxwell}, and dilaton are
\begin{subequations}
\begin{align}
F(z) &= (1-z)^2\sum_{n=0}^{\infty}{f_{n}(1-z)^n} \ ,\\
G(z) &= \frac{1}{(1-z)^2}\sum_{n=0}^{\infty}{g_{n}(1-z)^n} \ ,\\
A(z) &= \sum_{n=0}^{\infty}{a_{n}(1-z)^n} \ ,\\
h(z) &= h_{0}(1-z) \ ,\\
\Phi(z) & =\sum_{n=0}^{\infty}{\Phi_{n}(1-z)^n} \ .
\end{align}
\label{eqn:IRDDseries}
\end{subequations}
The equations of motion fix all coefficients except for $\{f_{0},a_{1},\Phi_{0}\}$, and we also have a choice of the value of the magnetic field $B$. We fix $f_{0}=1$ and generate different backgrounds that give different dual UV field theory parameters by choices of $B,a_{1}$ and $\Phi_{0}$. Our particular choice of ansatz actually fixes $Q=B$ for all of our solutions.\\
\indent In Figure~\ref{fig:EEstripDD}, we plot $s_{S}$ for the dilaton--dyon black hole for a few values of $B/\mu^2$. We find that increasing the value of $B/\mu^2$ appears to shift the entire entanglement entropy down. 
Zooming in on $s_{S}(\ell)$ (Figure~\ref{fig:EEstripDDzoomed}), we see that the behavior of the entanglement entropy appears to be linear in $\ell$, $s_{S}(\ell)\sim\ell$, with the slope increasing as $B/\mu^2$ is decreased.\\
\indent The results for the disk\footnote{\label{fn:DDdisk} The numerical results for the dilaton--dyon $s_{D}$ in Figure~\ref{fig:EEdiskDD} have very little numerical variance in comparison to those for the PED and PMD solutions. This is due to the fact that for the dyon solution the dilaton does not diverge in the IR.} entanglement entropy are shown in Figure~\ref{fig:EEdiskDD}, where we again found it useful to switch to the radial coordinate defined in \reef{eqn:radialcoord}, for three values of $B/\mu^2$. We find that the entanglement entropy, $\frac{s_{D}(\ell)}{\ell}$, appears to be linear in $\ell$, which suggest that $s_{D}(\ell)\sim\ell^2$, which would be a ``volume'' law rather than an area law for the entanglement entropy. As with the strip entanglement entropy, increasing $B/\mu^2$ shifts $s_{D}/\ell$ down but it also leads to a much more noticeable shift in the slope of the linear part of $s_{D}/\ell$. 
 
\begin{figure}[h!]
\centering
\includegraphics[width=0.6\textwidth]{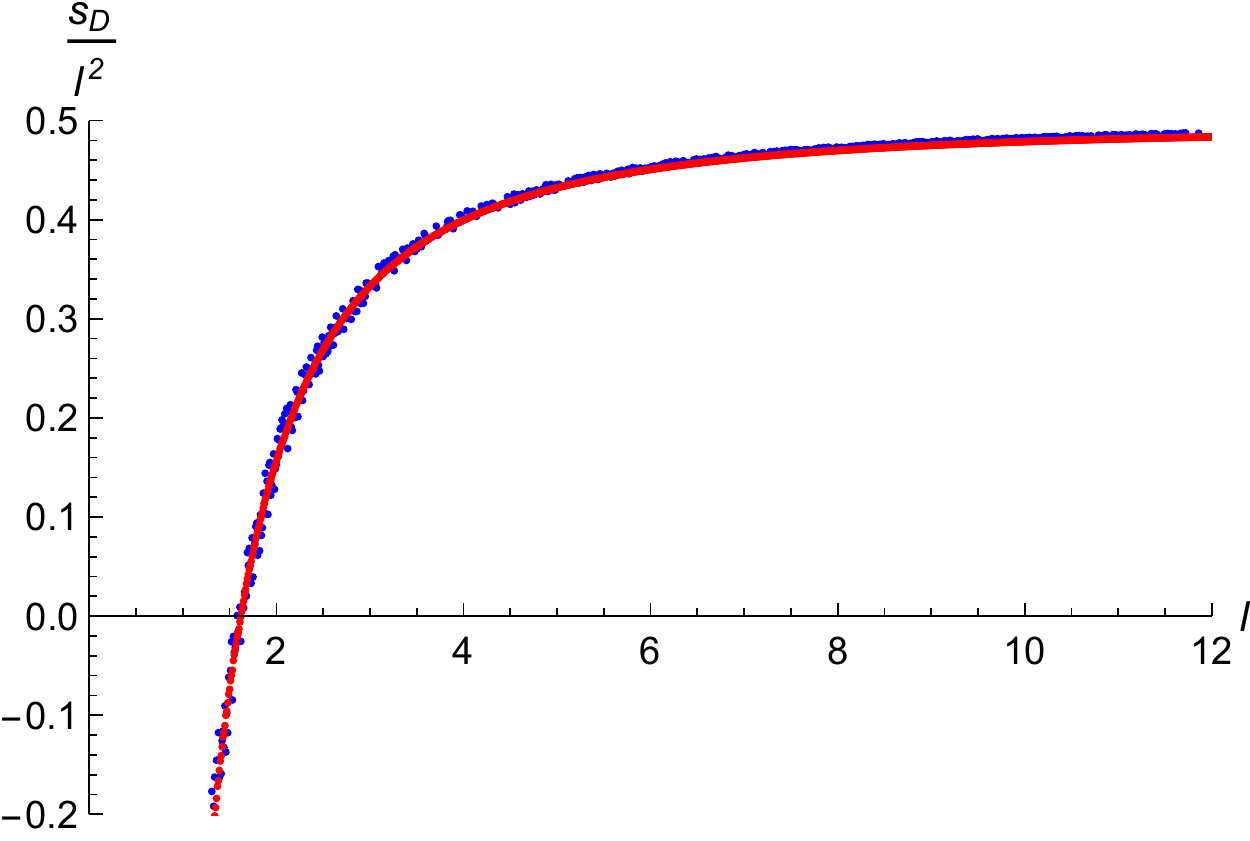}
\caption{Comparing the numerical and analytical dyonic black hole $s_{D}/\ell^2$. Blue is the analytical background and red is the numerical background.}
\label{fig:DDvsanalytic}
\end{figure}

 As a check, we turn off the dilaton (consistently setting $\Phi(z)\equiv 0$ in our equations of motion) and compute $s_{D}/\ell$ for the dyonic black hole. We can compare our numerical background's result to that of the analytical background written in ref.~\cite{Albash:2009wz}. The result is shown in Figure~\ref{fig:DDvsanalytic} and we see that there is very good agreement.
 
 All of our solutions are at zero temperature, and the PED black holes also have zero thermal entropy because they have zero area horizon, but our dilaton--dyon black hole has a finite area horizon (which in the dual gauge theory is interpreted as a large ground state degeneracy) and thus a finite entropy, $S$, even at zero temperature. Namely, computing the area of the horizon at $z=1$ using our IR expansion \reef{eqn:IRDDseries} (and dividing out by the infinite volume of $\mathbb{R}^2$), we find
\begin{equation}
\frac{S}{\text{Vol}(\mathbb{R}^2)}=\frac{\text{Area(Horizon)}}{4}=\frac{1}{4}\int{dxdy\sqrt{\frac{A(z)}{z^2}}}\Big|_{z=1}=\frac{\sqrt{a_{0}}}{4} \ ,
\end{equation}
which is a constant, where we have absorbed the factor of the AdS radius $L$ into the volume factor. In fact, from the equations of motion, we have that $a_{0}=\frac{B}{\sqrt{3}}$ and so
\begin{equation}
\frac{S}{\text{Vol}(\mathbb{R}^2)}=\frac{1}{3^{1/4}}\frac{B^{1/2}}{4} \ .
\label{eqn:groundstateentropy}
\end{equation}
 To see whether or not our entanglement entropy approaches the ground state entropy in the large~$\ell$ limit, we did a numerical fit to the linear part of $s_{D}/\ell$ in Figure~\ref{fig:EEdiskDDoverell}. The resulting slope is in Table \ref{table1}, comparing to the value of eqn.~\reef{eqn:groundstateentropy}  using the IR value of $B$.
 
\begin{table}
\centering
\begin{tabular}{| l | l | l | l | l |}
\hline
\text{Plot Color/Location} & \text{UV} $B/\mu^2$ & \text{IR} $B$ & $B^{1/2}/(3^{1/4}\cdot4)$ & \text{Slope} \\ \hline
\text{Blue/Upper} & 3.249 & 0.3074 & 0.1053 & 0.1011 \\ \hline
\text{Red/Middle} & 6.779 & 0.1477 & 0.0730 & 0.0569 \\ \hline
\text{Orange/Lower} & 27.866 & 0.035 & 0.0359 & 0.0322 \\
\hline
\end{tabular}
\caption{Comparing the numerically determined slope of the linear part of Figure~\ref{fig:EEdiskDDoverell} for $s_{D}/\ell$ to the ground state degeneracy in eqn.~\reef{eqn:groundstateentropy}. The plot colors/location and UV value of $B/\mu^2$  are given for ease of reference, as is the IR value of $B$, which is used to determine the ground state degeneracy via eqn.~\reef{eqn:groundstateentropy}.}
\label{table1}
\end{table}

\section{Entanglement Entropy of the Magnetic Electron Star} \label{MES}
\begin{figure}[t]
\centering
\subfigure[]{\includegraphics[width=3in]{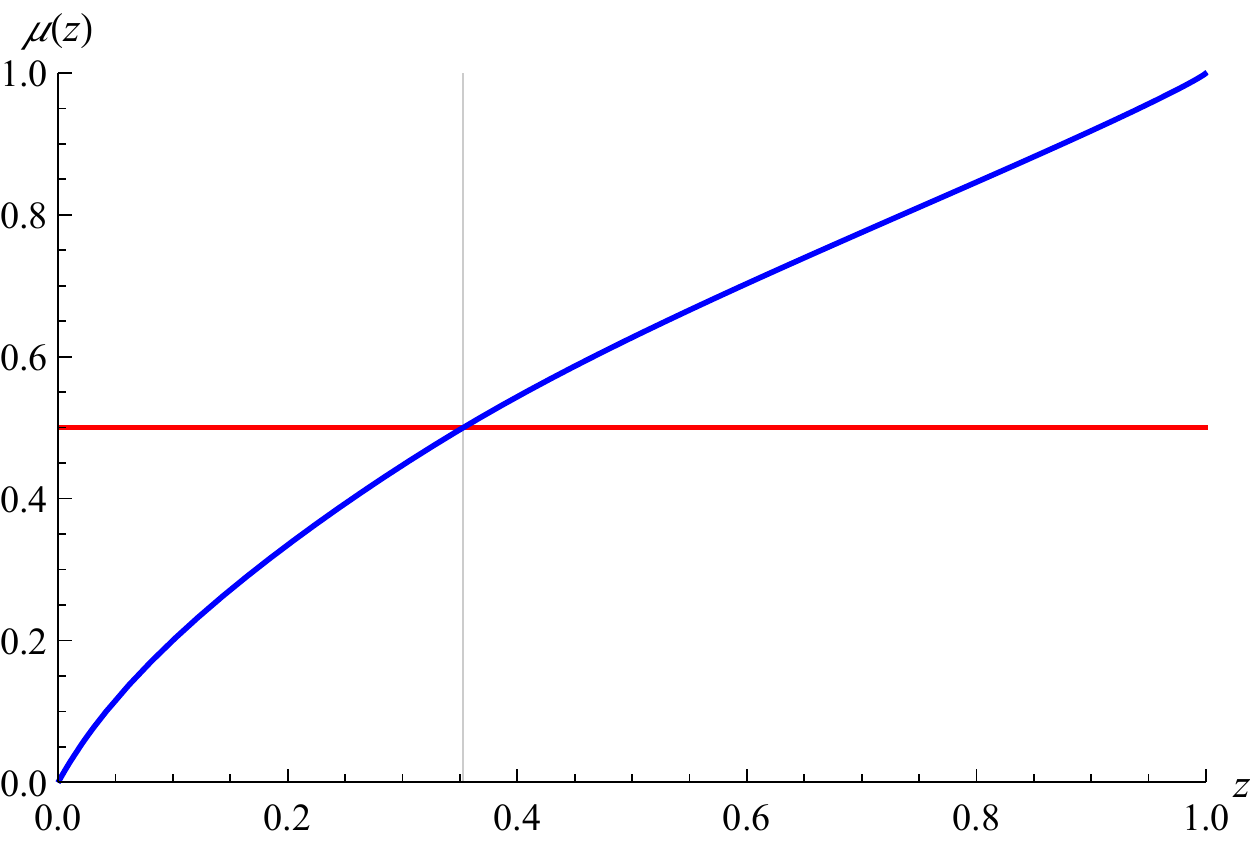}\label{fig:MESchemicalpotential}}
\subfigure[]{\includegraphics[width=3in]{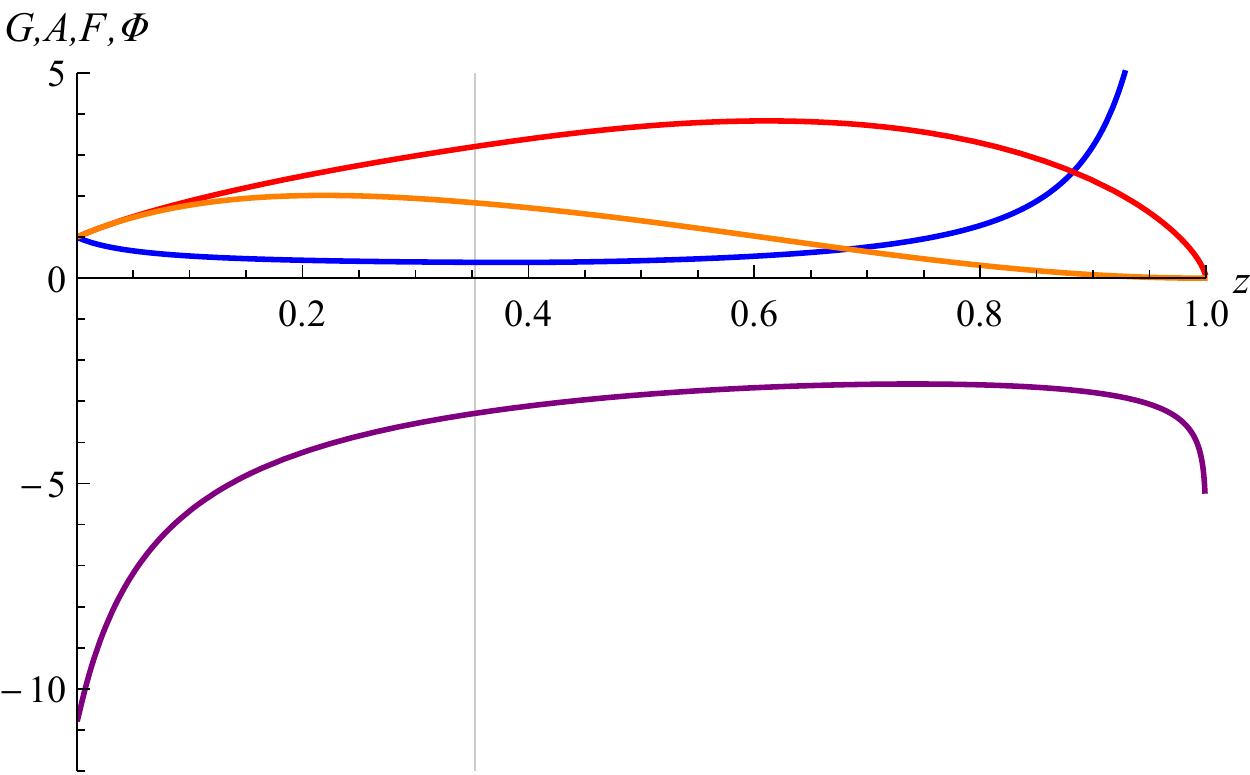}\label{fig:MESbackground}}
\caption{(a) The local chemical potential is shown in blue (the curve), where the fermion mass (shown in red, a horizontal line) is $\tilde{m}=0.5$ and $z_{\text{star}}=0.3529$ is the vertical line. The dual UV field theory parameters are $B/\mu^2=6.8132$, $Q/\mu^2=0.1136$, $\phi_1/\mu=-3.6507$, and $\phi_2/\mu^2=12.4852$. (b) The background metric and dilaton (lowest purple curve) for the same UV parameters as in (a). Here we have actually plotted $G(z)=z^2g(z), A(z)=z^2a(z)$, and $F(z)=z^2f(z)$ (lower--middle blue curve, upper red curve, and upper--middle orange curve respectively) so that our expect UV behavior is that each metric function goes to 1. We have also pulled out a factor of $z$ in $\Phi(z)$, redefining $\Phi(z) \rightarrow z\Phi(z)$ (purple). The vertical line is $z_{star}$.}
\end{figure}

We are now ready to turn on the star. The solutions were found in ref.~\cite{Albash:2012ht} in coordinates where the horizon is at infinity, but here we find it better to rescale those solutions so that the horizon is at $z=1$. We will consider the mesonic phase, where there is a charged star in the IR as well as a horizon, but with $\textit{no}$ electric charge behind the horizon. The IR expansion of the fields is given by
\begin{subequations}
\begin{align}
F(z) &= z^2(1-z)^{2}\left(\sum_{n=0}^{\infty}{f_{n}(1-z)^{2n/3}}+\delta f (1-z)^{2b/3}\right),\\
G(z) &= z^2(1-z)^{-4/3}\left(\sum_{n=0}^{\infty}{g_{n}(1-z)^{2n/3}}+\delta g (1-z)^{2b/3}\right),\\
A(z) &= z^2(1-z)^{2/3}\left(\sum_{n=0}^{\infty}{a_{n}(1-z)^{2n/3}}+\delta a (1-z)^{2b/3}\right),\\
h(z) &= h_{0}(1-z),\\
\Phi(z) &= \frac{\sqrt{3}}{3}\ln{(1-z)}+\sum_{n=0}^{\infty}{\Phi_{n}(1-z)^{2n/3}}+\delta\Phi (1-z)^{2b/3}.
\end{align}
\end{subequations}
We have again turned on a perturbation with $b = \frac{1}{3}\left(-3+\sqrt{57}\right)$ in order to flow to different values of the UV parameters \reef{eqn:dimensionlessratios}. The equations of motion fix all coefficients except $\{B,f_{0},\Phi_{0},\delta\Phi\}$. We choose a value of the fermion mass, $\tilde{m}$, and choose $\{B,f_{0},\delta\Phi\}$ and then adjust $\Phi_{0}$ (and rescale the time coordinate) so that the solution flows to $\text{AdS}_{4}$ in the UV. Figure~\ref{fig:MESchemicalpotential} shows an example, plotting the local chemical potential, fermion mass, and value of the ending radius of the star, $z_{star}$; Figure~\ref{fig:MESbackground} shows the metric functions and dilaton for the same background.

\begin{figure}[h!]
\centering
\subfigure[]{\includegraphics[width=3in]{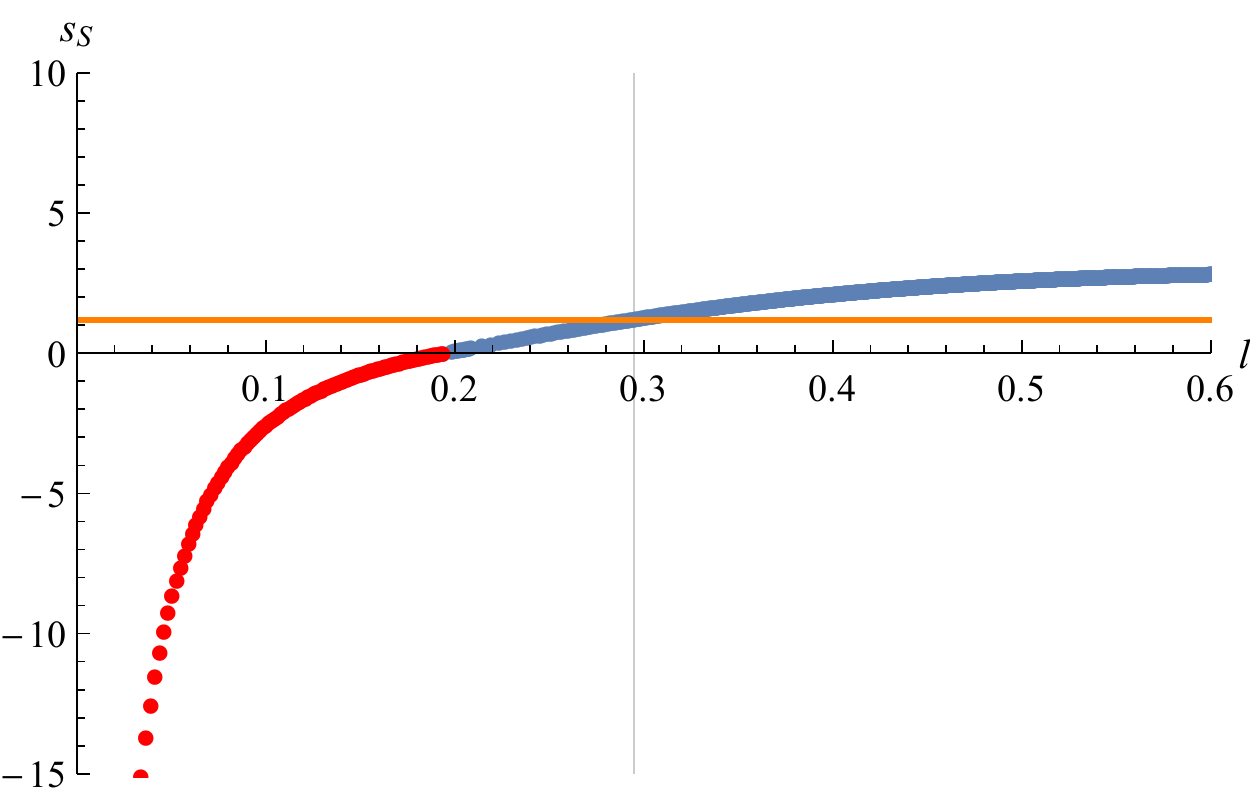}\label{fig:EEstripMES1}}
\subfigure[]{\includegraphics[width=3in]{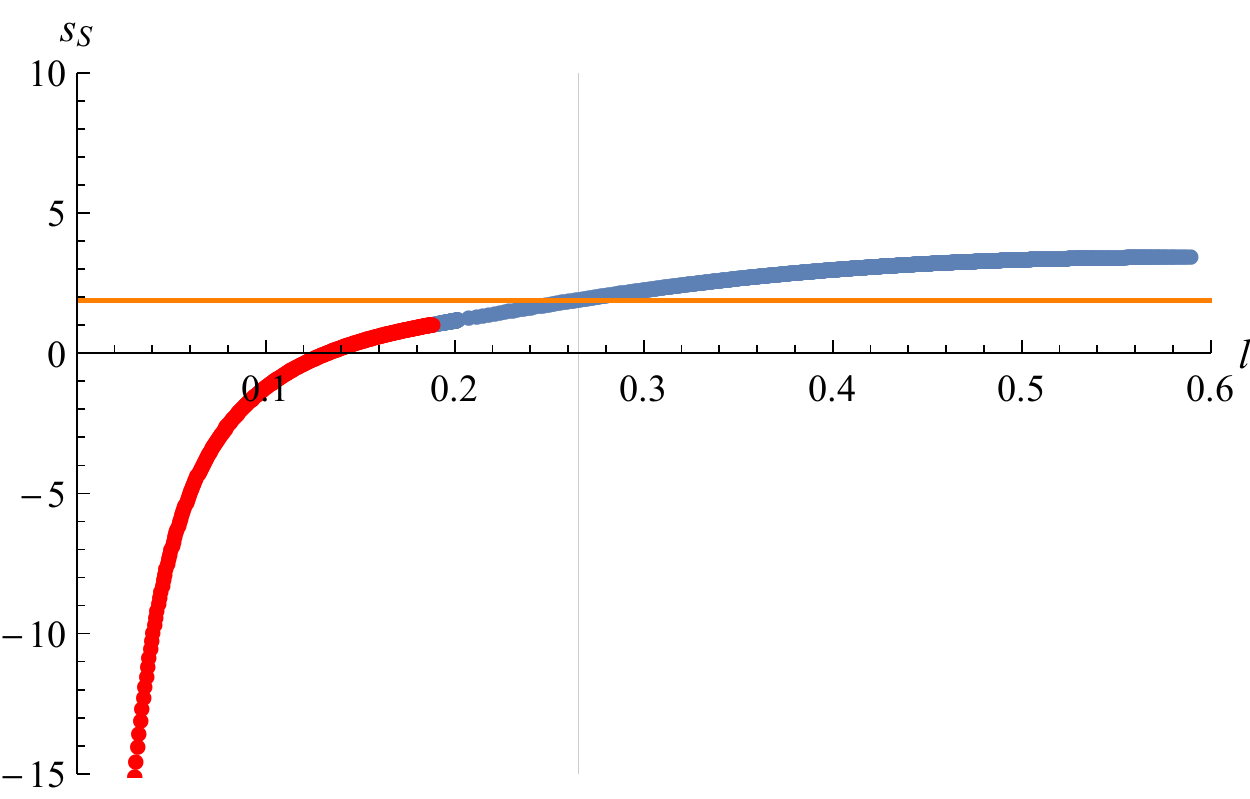}\label{fig:EEstripMES2}}
\caption{(a) The finite part of the entanglement entropy $s_{S}$ for the strip geometry as a function of the strip width $\ell$ for the mesonic phase of the magnetic electron star with $B/\mu^2=6.8132$. The red, lower part of the curve (that tends to negative infinity), is the part of $s_{S}$ that lies outside the star and the blue, upper part of the curve (larger values of $\ell$), is the part that lies inside the star; the minimal surface crosses into the star at a value of $\ell_{\text{star}}=0.1938$. The orange, horizontal line is the value of $s_{S}$ for the ``in--falling'' solution and the intersection of the horizontal and vertical lines indicate the point at which the entanglement entropy crosses over to the in--falling one. (b) Same as (a) but for $B/\mu^2=14.1343$. Here the minimal surface crosses into the star at a value of $\ell_{\text{star}}=0.1883$. This differs from (a) since the value of $z_{\text{star}}$ is different.}
\label{fig:EEstripMES}
\end{figure}

 For this background, we show in Figure~\ref{fig:EEstripMES1} the finite entanglement entropy for the strip geometry. As with the PED black hole solution, we have computed the area of the ``in--falling'' solution. We find that the large $\ell$ value of $s_{S}$ is greater than the area of the ``in--falling'' solution. In fact, we are able to generate several magnetic electron star backgrounds (see Figure~\ref{fig:MESphasediagram}) and from those backgrounds there are strong indications that for every one the large $\ell$ value of $s_{S}$ is greater than the in--falling solution, with the difference between the two always around $1.5\pm0.2$. However, we do not show curves for $s_{S}$ for all of these backgrounds because the data are too difficult to control numerically well enough to display -- there is too much numerical noise; nevertheless, the relationship between the in--falling and large $\ell$ value of $s_{S}$ is clear to us from these investigations. We also checked this behavior in a different coordinate system (one in which the horizon is at $z\rightarrow\infty$) and again find the same behavior. This suggests that the behavior is robust. Since the in--falling solution has a lower entropy than the minimal surface, we conclude that the actual entanglement entropy as seen in Figure~\ref{fig:EEstripMES} follows the red curve, then the blue, until it meets the orange line, and then it remains that of the in--falling solution for all $\ell$ thereafter (that is, it follows the curve starting from negative infinity all the way until it means the intersection of the vertical and horizontal lines, after which it remains the horizontal line). From the point of view of the minimal surface, at a certain value of $\ell$, the hanging surface breaks and becomes the in--falling solution. Figure~\ref{fig:EEstripMES2} shows the strip entanglement entropy for a different value of the dimensionless ratio~$B/\mu^2$ to demonstrate this point. We also find evidence that for the backgrounds generated along the line in Figure~\ref{fig:MESphasediagram} that the entanglement entropy increases as $B/\mu^2$ is increased (as explained above, we choose not to display this data due to the difficult numerics involved). 

\begin{figure}[h!]
\centering
\subfigure[]{\includegraphics[width=3in]{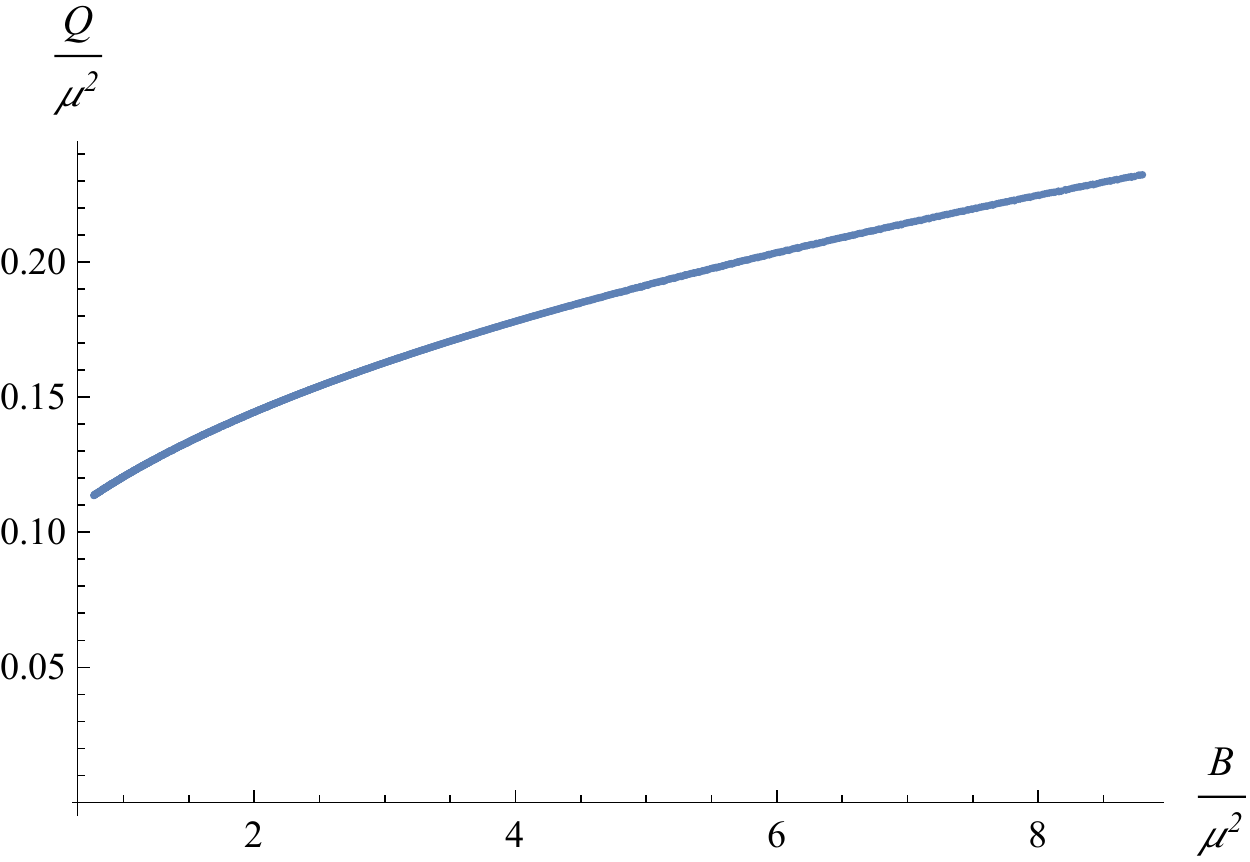}\label{fig:MESphasediagram1}}
\subfigure[]{\includegraphics[width=3in]{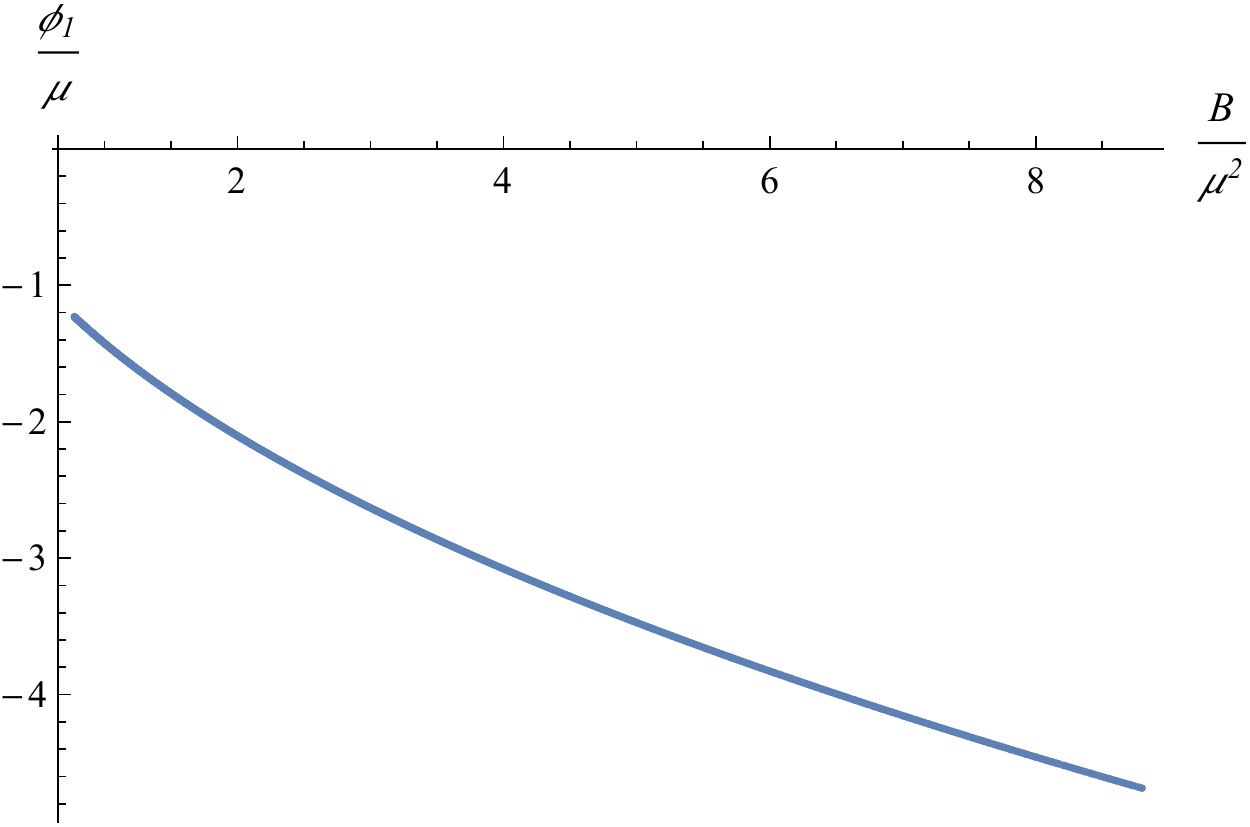}\label{fig:MESphasediagram}}
\caption{(a) The line in the phase diagram for the different magnetic electron star backgrounds we are able to generate, showing here the relationship between $B/\mu^2$ vs. $Q/\mu^2$, where $B/\mu^2$ is the free parameter. (b) The same line in the phase diagram but showing two of the parameters we control, $\phi_{1}/\mu$ vs. $B/\mu^2$.}
\label{fig:MESphasediagram}
\end{figure}

 We note that we were unable to compute the disk entanglement entropy for the mesonic phase of the magnetic electron star at this point in time due to numerical instability in solving the second order problem for these backgrounds, at least at the level of the numerical sophistication that we are using. In addition, although we have solutions for partially fractionalized phases of magnetic electron stars, given in ref. \cite{Albash:2012ht}, we are currently unable to change how populated the star is (i.e. how much the chemical potential is greater than the fermion mass) and so cannot fully explore the entanglement entropy of these solutions.

Our results strongly indicate that the entanglement entropy of the mesonic phase of the magnetic electron star is undergoing a ``phase transition'' as it transitions to the in--falling solution. This is reminiscent of the breaking of the hanging string in confinement/deconfinement phase transitions \cite{Maldacena:1998im,Rey:1998ik,Witten:1998zw}, and similar behavior has been seen in the work of two of the authors of this paper, ref. \cite{Albash:2012pd}, in the case of holographic superconductors. However, without being able to compare to the disk case we cannot say how robust this ``phase transition'' may be, and in addition, without a careful study of the stability of magnetic electron stars we cannot rule out the possibility that we are detecting an instability. These are matters that demand further investigation.

\section{Conclusion} \label{Conclusion}
We have computed the finite part of the holographic entanglement entropy for the strip and disk geometry for purely electric, purely magnetic, and dyonic dilatonic black holes, as well as the strip entanglement entropy for the magnetic electron star solutions of ref.~\cite{Albash:2012ht}. We observed how the entanglement entropy for the PED solution backgrounds at large $\ell$ approach the in--falling solution, which contributes a non--zero value to the entanglement entropy. We also found that different UV regulators, due to having different coordinate systems, can introduce a shift in the entanglement entropy. Similar results are obtained for the PMD backgrounds, and we are able to show that the entanglement entropy is invariant under electromagnetic duality, as long as one shifts the relative location of the physical horizons between the PED and PMD solutions.

 For the dilaton--dyon black hole, we found that the entanglement entropy for the disk, at large $\ell$, grows linearly with $\ell$, with a coefficient approximately equal to the ground state degeneracy of our solutions. This suggests a volume law rather than an area law for these backgrounds. Lastly, for the mesonic phase of the magnetic electron star solution, we found that the entanglement entropy smoothly passes into the star, but that at a certain value of $\ell$, it is actually the in--falling solution that has lower entropy than the hanging surface. This suggests that at a value of $z$, with $z_{star}<z<1$, the minimal surface breaks to the in--falling one. This strongly suggests the presence of a ``phase transition,'' however, pending further investigation we cannot say for certain; regardless, our results indicate at least some non--trivial behavior in the far IR for the entanglement entropy. In all cases, we find a dependence of the large $\ell$ behavior of the entanglement entropy on the various dimensionless ratios in the theory.
 
 There are several directions for future work. We mentioned in Section \ref{MES} that one could study the disk entanglement entropy for the mesonic case, as well as the entanglement entropy of the partially fractionalized phases of the magnetic electron stars.  For the case of zero magnetic field but with a charged star and electrically charged horizon--a partially fractionalized phase--the entanglement entropy's dependence on the charge was studied in refs. \cite{Huijse:2011ef} and there it was found that there was a logarithmic violation to the area law due to the charged horizon but not the star. It would be interesting to see if the presence of the magnetic field in such phases changes this behavior. Since we have only studied the mesonic phase in this paper, we do not expect the logarithmic violation, at least in the electric sector. Perhaps, however, even in the mesonic phase the presence of the horizon for the magnetic sector will modify the area law behavior of the entanglement entropy in an interesting way. Additionally, one could construct and study finite temperature versions of our solutions. These points are left for future work.

\section{Acknowledgments}
SM and CVJ would like to thank the US Department of Energy for support under grant DE-FG03-84ER-40168. SM thanks the USC Dornsife College of Letters, Arts, and Sciences, and Jessica Chen for her help with graphics, coding and support.

\section*{Appendix}
 We present here the details of the derivations of eqns. \reef{eqn:EEstripintegral}, \reef{eqn:EEstriplength}, and \reef{eqn:EEdiskpullback} given in Section \ref{EEReview}. Recall we have taken bulk metric to be of the form
\begin{equation}
ds^2=-g_{tt}(z)dt^2+g_{zz}(z)dz^2+g_{xx}(z)d\vec{x}^2 \ ,
\end{equation}
where UV is at $z=0$ and the horizon at $z=1$, after rescaling the coordinates by $z_{H}$, the physical horizon position. Writing the $\mathbb{R}^2$ part of the metric as $dx^2+dy^2$ and having the finite length of the strip run from $-\ell/2 \leq x\leq \ell/2$, the minimal surface will be symmetric about $y=0$. We denote the~$z$ value of the turning point of the surface as $z_{T}$. Let the surface have coordinates $(x,y)$, the same as the $\mathbb{R}^2$ coordinates, with embedding $z=z(x)$. Then we may write the pull--back of the metric, at fixed time, as 
\begin{equation}
ds^{2}_{\gamma_{A}}= \left(g_{zz}(z(x))\left(\frac{dz}{dx}\right)^2+g_{xx}(z(x))\right)dx^2+g_{xx}(z(x))dy^2 \ ,
\end{equation}
and thus the finite part of the entanglement entropy becomes
\begin{equation}
4 G_{N}S_{S} =2\mathscr{L}\int_{0}^{\ell/2}{dx g_{xx}(z(x))\sqrt{1+\frac{g_{zz}(z(x))}{g_{xx}(z(x))}\left(\frac{dz}{dx}\right)^2}} \ .
\label{Aeqn:strippullback}
\end{equation}
Because the above equation has no explicit dependence on the coordinate $x$, we can reduce the second order problem of minimizing this integral for the surface $z(x)$ into a first order problem~\cite{Ryu:2006ef}. We see that eqn.~\reef{Aeqn:strippullback} does not have any explicit dependence on $x$, so thinking of it as a time coordinate the associated ``Hamiltonian" will be a conserved quantity with respect to $x$, i.e., $\frac{dH}{dx}=0$. We denote by $p_{z}$ the conjugate momentum for the variable $z$, and taking the integrand in eqn.~\reef{Aeqn:strippullback} as the ``Lagrangian", we find
\begin{equation}
p_{z}=\frac{g_{zz}(z(x))z'(x)}{\sqrt{1+\frac{g_{zz}(z(x))}{g_{xx}(z(x))}\left(z'(x)\right)^2}} \ ,
\end{equation}
which allows us to write the conserved ``Hamiltonian''
\begin{equation}
H=-\frac{g_{xx}(z(x))}{\sqrt{1+\frac{g_{zz}(z(x))}{g_{xx}(z(x))}\left(z'(x)\right)^2}}=\text{constant}.
\label{Aeqn:Hamiltonian}
\end{equation}
This constant of motion is related to the turning point $z_{T}$, as we will shortly see, so we denote the constant by $c(z_{T})$ and, squaring both sides of eqn.~\reef{Aeqn:Hamiltonian} and doing some algebra, we find
\begin{equation}
\frac{dz}{dx}=\sqrt{\frac{1}{c^2(z_{T})}\frac{g_{xx}(z)}{g_{zz}(z)}\left(g_{xx}^2(z)-c^2(z_{T})\right)} \ .
\label{Aeqn:firstorderproblem}
\end{equation} 
This is our first integral of motion, reducing the original second order problem of finding $z(x)$ to this first order one. We can go further, however, and bypass solving for $z(x)$ directly. First, from eqn.~\reef{Aeqn:firstorderproblem} we see that the constant $c(z_{T})$ is related to the turning point if we set
\begin{equation}
c(z_{T})=\pm g_{xx}(z_{T}) \ .
\end{equation}
Then we have that 
\begin{equation}
dx=\frac{g_{xx}(z_{T})dz}{\sqrt{\frac{g_{xx}(z)}{g_{zz}(z)}\left(g_{xx}^2(z)-g_{xx}^2(z_{T})\right)}} \ .
\end{equation}
This last relationship allows us to write the original entanglement entropy integral eqn.~\reef{Aeqn:strippullback} as an integral over $z$. The result is
\begin{equation}
4 G_{N}S_{S} =2\mathscr{L}\int_{\epsilon}^{z_{T}}{dz \frac{g_{xx}(z)^2}{\sqrt{\frac{g_{xx}(z)}{g_{zz}(z)}\left(g_{xx}(z)^2-g_{xx}(z_{T})^2\right)}}} \ ,
\label{Aeqn:EEstripintegral}
\end{equation}
where we have introduced the UV cutoff $\epsilon$ in the lower limit of the integral. With this, we can easily find the length of the strip, $\ell=2\int_{0}^{\ell/2}{dx}$, as a function of the turning point via substitution of $dx$ above; we find
\begin{equation}
\frac{\ell}{2}=\int_{0}^{z_{T}}{dz\frac{g_{xx}(z_{T})}{\sqrt{\frac{g_{xx}(z)}{g_{zz}(z)}\left(g_{xx}(z)^2-g_{xx}(z_{T})^2\right)}}} \ .
\label{Aeqn:EEstriplength}
\end{equation}
Since the length $\ell$ should be positive, in writing eqn.~\reef{Aeqn:EEstriplength} we made the choice $c(z_{T})=-g_{xx}(z_{T})$.

 The same trick does not work, however, for the disk geometry, and so the formula is simply given by the pull--back
\begin{equation}
4 G_{N}S_{D} =2\pi\int_{0}^{\ell}{dr\, r g_{xx}(z(r))\sqrt{1+\frac{g_{zz}(z(r))}{g_{xx}(z(r))}\left(\frac{dz}{dr}\right)^2}} \ ,
\label{Aeqn:EEdiskpullback}
\end{equation}
where $0\leq r\leq \ell$ is the radial variable for polar coordinates of the $\mathbb{R}^2$, $dr^2+r^2d\theta^2$. In this case, we must explicitly solve for the minimal surface $z(r)$ and then input that into eqn.~\reef{Aeqn:EEdiskpullback} to find the entanglement entropy. Eqns. \reef{Aeqn:EEstripintegral}, \reef{Aeqn:EEstriplength}, and \reef{Aeqn:EEdiskpullback} are the same as eqns. \reef{eqn:EEstripintegral}, \reef{eqn:EEstriplength}, and \reef{eqn:EEdiskpullback} given in Section \ref{EEReview}.
									

\begin{thebibliography}{10}

\bibitem{Maldacena:1997re}
J.~M. Maldacena, ``{The Large N limit of superconformal field theories and
  supergravity},'' \href{http://dx.doi.org/10.1023/A:1026654312961}{{\em
  Int.J.Theor.Phys.} {\bfseries 38} (1999) 1113--1133},
\href{http://arxiv.org/abs/hep-th/9711200}{{\ttfamily arXiv:hep-th/9711200
  [hep-th]}}.

\bibitem{Witten:1998qj}
E.~Witten, ``Anti-de sitter space and holography,'' {\em Adv. Theor. Math.
  Phys.} {\bfseries 2} (1998) 253--291,
\href{http://arxiv.org/abs/hep-th/9802150}{{\ttfamily hep-th/9802150}}.

\bibitem{Gubser:1998bc}
S.~S. Gubser, I.~R. Klebanov, and A.~M. Polyakov, ``Gauge theory correlators
  from non-critical string theory,'' {\em Phys. Lett.} {\bfseries B428} (1998)
  105--114,
\href{http://arxiv.org/abs/hep-th/9802109}{{\ttfamily hep-th/9802109}}.

\bibitem{Aharony:1999t}
O.~Aharony, S.~S. Gubser, J.~M. Maldacena, H.~Ooguri, and Y.~Oz, ``Large n
  field theories, string theory and gravity,'' {\em Phys. Rept.} {\bfseries
  323} (2000) 183--386,
\href{http://arxiv.org/abs/hep-th/9905111}{{\ttfamily hep-th/9905111}}.

\bibitem{Kovtun:2004de}
P.~Kovtun, D.~T. Son, and A.~O. Starinets, ``Viscosity in strongly interacting
  quantum field theories from black hole physics,'' {\em Phys. Rev. Lett.}
  {\bfseries 94} (2005) 111601,
\href{http://arxiv.org/abs/hep-th/0405231}{{\ttfamily hep-th/0405231}}.

\bibitem{Iqbal:2008by}
N.~Iqbal and H.~Liu, ``{Universality of the hydrodynamic limit in AdS/CFT and
  the membrane paradigm},''
  \href{http://dx.doi.org/10.1103/PhysRevD.79.025023}{{\em Phys.Rev.}
  {\bfseries D79} (2009) 025023},
\href{http://arxiv.org/abs/0809.3808}{{\ttfamily arXiv:0809.3808 [hep-th]}}.

\bibitem{Witten:1998zw}
E.~Witten, ``Anti-de sitter space, thermal phase transition, and confinement in
  gauge theories,'' {\em Adv. Theor. Math. Phys.} {\bfseries 2} (1998)
  505--532,
\href{http://arxiv.org/abs/hep-th/9803131}{{\ttfamily hep-th/9803131}}.

\bibitem{McGreevy:2009xe}
J.~McGreevy, ``{Holographic duality with a view toward many-body physics},''
  \href{http://dx.doi.org/10.1155/2010/723105}{{\em Adv.High Energy Phys.}
  {\bfseries 2010} (2010) 723105},
\href{http://arxiv.org/abs/0909.0518}{{\ttfamily arXiv:0909.0518 [hep-th]}}.

\bibitem{Hartnoll:2009sz}
S.~A. Hartnoll, ``{Lectures on holographic methods for condensed matter
  physics},''
\href{http://arxiv.org/abs/0903.3246}{{\ttfamily arXiv:0903.3246 [hep-th]}}.

\bibitem{Hartnoll:2011fn}
S.~A. Hartnoll, ``{Horizons, holography and condensed matter},''
\href{http://arxiv.org/abs/1106.4324}{{\ttfamily arXiv:1106.4324 [hep-th]}}.

\bibitem{Ogawa:2011bz}
N.~Ogawa, T.~Takayanagi, and T.~Ugajin, ``{Holographic Fermi Surfaces and
  Entanglement Entropy},''
  \href{http://dx.doi.org/10.1007/JHEP01(2012)125}{{\em JHEP} {\bfseries 1201}
  (2012) 125},
\href{http://arxiv.org/abs/1111.1023}{{\ttfamily arXiv:1111.1023 [hep-th]}}.

\bibitem{Huijse:2011ef}
L.~Huijse, S.~Sachdev, and B.~Swingle, ``{Hidden Fermi surfaces in compressible
  states of gauge-gravity duality},''
  \href{http://dx.doi.org/10.1103/PhysRevB.85.035121}{{\em Phys.Rev.}
  {\bfseries B85} (2012) 035121},
\href{http://arxiv.org/abs/1112.0573}{{\ttfamily arXiv:1112.0573
  [cond-mat.str-el]}}.

\bibitem{Sachdev:2010um}
S.~Sachdev, ``{Holographic metals and the fractionalized Fermi liquid},''
  \href{http://dx.doi.org/10.1103/PhysRevLett.105.151602}{{\em Phys.Rev.Lett.}
  {\bfseries 105} (2010) 151602},
\href{http://arxiv.org/abs/1006.3794}{{\ttfamily arXiv:1006.3794 [hep-th]}}.

\bibitem{Huijse:2011hp}
L.~Huijse and S.~Sachdev, ``{Fermi surfaces and gauge-gravity duality},''
  \href{http://dx.doi.org/10.1103/PhysRevD.84.026001}{{\em Phys.Rev.}
  {\bfseries D84} (2011) 026001},
\href{http://arxiv.org/abs/1104.5022}{{\ttfamily arXiv:1104.5022 [hep-th]}}.

\bibitem{Lee:2008xf}
S.-S. Lee, ``{A Non-Fermi Liquid from a Charged Black Hole: A Critical Fermi
  Ball},'' \href{http://dx.doi.org/10.1103/PhysRevD.79.086006}{{\em Phys. Rev.}
  {\bfseries D79} (2009) 086006},
\href{http://arxiv.org/abs/0809.3402}{{\ttfamily arXiv:0809.3402 [hep-th]}}.

\bibitem{Liu:2009dm}
H.~Liu, J.~McGreevy, and D.~Vegh, ``{Non-Fermi liquids from holography},''
\href{http://arxiv.org/abs/0903.2477}{{\ttfamily arXiv:0903.2477 [hep-th]}}.

\bibitem{Cubrovic:2009ye}
M.~Cubrovic, J.~Zaanen, and K.~Schalm, ``{Fermions and the AdS/CFT
  correspondence: quantum phase transitions and the emergent Fermi-liquid},''
\href{http://arxiv.org/abs/0904.1993}{{\ttfamily arXiv:0904.1993 [hep-th]}}.

\bibitem{Faulkner:2009wj}
T.~Faulkner, H.~Liu, J.~McGreevy, and D.~Vegh, ``{Emergent quantum criticality,
  Fermi surfaces, and AdS2},''
\href{http://arxiv.org/abs/0907.2694}{{\ttfamily arXiv:0907.2694 [hep-th]}}.

\bibitem{Hartnoll:2010xj}
S.~A. Hartnoll, D.~M. Hofman, and A.~Tavanfar, ``{Holographically smeared Fermi
  surface: Quantum oscillations and Luttinger count in electron stars},''
  \href{http://dx.doi.org/10.1209/0295-5075/95/31002}{{\em Europhys.Lett.}
  {\bfseries 95} (2011) 31002},
\href{http://arxiv.org/abs/1011.2502}{{\ttfamily arXiv:1011.2502 [hep-th]}}.

\bibitem{Hartnoll:2011dm}
S.~A. Hartnoll, D.~M. Hofman, and D.~Vegh, ``{Stellar spectroscopy: Fermions
  and holographic Lifshitz criticality},'' {\em JHEP} {\bfseries 1108} (2011)
  096,
\href{http://arxiv.org/abs/1105.3197}{{\ttfamily arXiv:1105.3197 [hep-th]}}.

\bibitem{Hartnoll:2010gu}
S.~A. Hartnoll and A.~Tavanfar, ``{Electron stars for holographic metallic
  criticality},'' \href{http://dx.doi.org/10.1103/PhysRevD.83.046003}{{\em
  Phys.Rev.} {\bfseries D83} (2011) 046003},
\href{http://arxiv.org/abs/1008.2828}{{\ttfamily arXiv:1008.2828 [hep-th]}}.

\bibitem{Iqbal:2011in}
N.~Iqbal, H.~Liu, and M.~Mezei, ``{Semi-local quantum liquids},'' {\em JHEP}
  {\bfseries 1204} (2012) 086,
\href{http://arxiv.org/abs/1105.4621}{{\ttfamily arXiv:1105.4621 [hep-th]}}.

\bibitem{Sachdev:2011ze}
S.~Sachdev, ``{A model of a Fermi liquid using gauge-gravity duality},'' {\em
  Phys.Rev.} {\bfseries D84} (2011) 066009,
\href{http://arxiv.org/abs/1107.5321}{{\ttfamily arXiv:1107.5321 [hep-th]}}.

\bibitem{Hartnoll:2011pp}
S.~A. Hartnoll and L.~Huijse, ``{Fractionalization of holographic Fermi
  surfaces},'' \href{http://dx.doi.org/10.1088/0264-9381/29/19/194001}{{\em
  Class.Quant.Grav.} {\bfseries 29} (2012) 194001},
\href{http://arxiv.org/abs/1111.2606}{{\ttfamily arXiv:1111.2606 [hep-th]}}.

\bibitem{Ryu:2006bv}
S.~Ryu and T.~Takayanagi, ``{Holographic derivation of entanglement entropy
  from AdS/CFT},'' \href{http://dx.doi.org/10.1103/PhysRevLett.96.181602}{{\em
  Phys.Rev.Lett.} {\bfseries 96} (2006) 181602},
\href{http://arxiv.org/abs/hep-th/0603001}{{\ttfamily arXiv:hep-th/0603001
  [hep-th]}}.

\bibitem{Albash:2012ht}
T.~Albash, C.~V. Johnson, and S.~MacDonald, ``{Holography, Fractionalization
  and Magnetic Fields},''
  \href{http://dx.doi.org/10.1007/978-3-642-37305-3\_20}{{\em Lect.Notes Phys.}
  {\bfseries 871} (2013) 537--554},
\href{http://arxiv.org/abs/1207.1677}{{\ttfamily arXiv:1207.1677 [hep-th]}}.

\bibitem{Kundu:2012jn}
N.~Kundu, P.~Narayan, N.~Sircar, and S.~P. Trivedi, ``{Entangled Dilaton
  Dyons},'' \href{http://dx.doi.org/10.1007/JHEP03(2013)155}{{\em JHEP}
  {\bfseries JHEP03} (2013) 155},
\href{http://arxiv.org/abs/1208.2008}{{\ttfamily arXiv:1208.2008 [hep-th]}}.

\bibitem{Iizuka:2011hg}
N.~Iizuka, N.~Kundu, P.~Narayan, and S.~P. Trivedi, ``{Holographic Fermi and
  Non-Fermi Liquids with Transitions in Dilaton Gravity},''
  \href{http://dx.doi.org/10.1007/JHEP01(2012)094}{{\em JHEP} {\bfseries 1201}
  (2012) 094},
\href{http://arxiv.org/abs/1105.1162}{{\ttfamily arXiv:1105.1162 [hep-th]}}.

\bibitem{Carney:2015dra}
D.~Carney and M.~Edalati, ``{Dyonic Stars for Holography},''
\href{http://arxiv.org/abs/1501.06938}{{\ttfamily arXiv:1501.06938 [hep-th]}}.

\bibitem{Puletti:2015gwa}
V.~G.~M. Puletti, S.~Nowling, L.~Thorlacius, and T.~Zingg, ``{Magnetic
  oscillations in a holographic liquid},''
\href{http://arxiv.org/abs/1501.06459}{{\ttfamily arXiv:1501.06459 [hep-th]}}.

\bibitem{Ryu:2006ef}
S.~Ryu and T.~Takayanagi, ``{Aspects of Holographic Entanglement Entropy},''
  \href{http://dx.doi.org/10.1088/1126-6708/2006/08/045}{{\em JHEP} {\bfseries
  0608} (2006) 045},
\href{http://arxiv.org/abs/hep-th/0605073}{{\ttfamily arXiv:hep-th/0605073
  [hep-th]}}.

\bibitem{deRitis:1985uw}
R.~de~Ritis, M.~Lavorgna, G.~Platania, and C.~Stornaiolo, ``{Charged spin fluid
  in the Einstein-Cartan theory},''
\href{http://dx.doi.org/10.1103/PhysRevD.31.1854}{{\em Phys.Rev.} {\bfseries
  D31} (1985) 1854--1859}.

\bibitem{Bombelli:1990ze}
L.~Bombelli and R.~Torrence, ``{Perfect fluids and Ashtekar variables, with
  applications to Kantowski-Sachs models},''
\href{http://dx.doi.org/10.1088/0264-9381/7/10/008}{{\em Class.Quant.Grav.}
  {\bfseries 7} (1990) 1747--1765}.

\bibitem{Brown:1992kc}
J.~D. Brown, ``{Action functionals for relativistic perfect fluids},''
  \href{http://dx.doi.org/10.1088/0264-9381/10/8/017}{{\em Class.Quant.Grav.}
  {\bfseries 10} (1993) 1579--1606},
\href{http://arxiv.org/abs/gr-qc/9304026}{{\ttfamily arXiv:gr-qc/9304026
  [gr-qc]}}.

\bibitem{Gubser:2009qt}
S.~S. Gubser and F.~D. Rocha, ``{Peculiar properties of a charged dilatonic
  black hole in ${\text{AdS}_5}$},''
  \href{http://dx.doi.org/10.1103/PhysRevD.81.046001}{{\em Phys.Rev.}
  {\bfseries D81} (2010) 046001},
\href{http://arxiv.org/abs/0911.2898}{{\ttfamily arXiv:0911.2898 [hep-th]}}.

\bibitem{Albash:2011nq}
T.~Albash and C.~V. Johnson, ``{Holographic Entanglement Entropy and
  Renormalization Group Flow},''
  \href{http://dx.doi.org/10.1007/JHEP02(2012)095}{{\em JHEP} {\bfseries 1202}
  (2012) 095},
\href{http://arxiv.org/abs/1110.1074}{{\ttfamily arXiv:1110.1074 [hep-th]}}.

\bibitem{Witten:2003ya}
E.~Witten, ``{SL(2,Z) action on three-dimensional conformal field theories with
  Abelian symmetry},''
\href{http://arxiv.org/abs/hep-th/0307041}{{\ttfamily arXiv:hep-th/0307041
  [hep-th]}}.

\bibitem{Hartnoll:2007ih}
S.~A. Hartnoll, P.~K. Kovtun, M.~Muller, and S.~Sachdev, ``{Theory of the
  Nernst effect near quantum phase transitions in condensed matter, and in
  dyonic black holes},''
  \href{http://dx.doi.org/10.1103/PhysRevB.76.144502}{{\em Phys.Rev.}
  {\bfseries B76} (2007) 144502},
\href{http://arxiv.org/abs/0706.3215}{{\ttfamily arXiv:0706.3215
  [cond-mat.str-el]}}.

\bibitem{Albash:2009wz}
T.~Albash and C.~V. Johnson, ``{Holographic Aspects of Fermi Liquids in a
  Background Magnetic Field},''
  \href{http://dx.doi.org/10.1088/1751-8113/43/34/345405}{{\em J.Phys.}
  {\bfseries A43} (2010) 345405},
\href{http://arxiv.org/abs/0907.5406}{{\ttfamily arXiv:0907.5406 [hep-th]}}.

\bibitem{Maldacena:1998im}
J.~M. Maldacena, ``{Wilson loops in large N field theories},''
  \href{http://dx.doi.org/10.1103/PhysRevLett.80.4859}{{\em Phys.Rev.Lett.}
  {\bfseries 80} (1998) 4859--4862},
\href{http://arxiv.org/abs/hep-th/9803002}{{\ttfamily arXiv:hep-th/9803002
  [hep-th]}}.

\bibitem{Rey:1998ik}
S.-J. Rey and J.-T. Yee, ``Macroscopic strings as heavy quarks in large n gauge
  theory and anti-de sitter supergravity,'' {\em Eur. Phys. J.} {\bfseries C22}
  (2001) 379--394,
\href{http://arxiv.org/abs/hep-th/9803001}{{\ttfamily hep-th/9803001}}.

\bibitem{Albash:2012pd}
T.~Albash and C.~V. Johnson, ``{Holographic Studies of Entanglement Entropy in
  Superconductors},'' \href{http://dx.doi.org/10.1007/JHEP05(2012)079}{{\em
  JHEP} {\bfseries 1205} (2012) 079},
\href{http://arxiv.org/abs/1202.2605}{{\ttfamily arXiv:1202.2605 [hep-th]}}.

\end{thebibliography}

\providecommand{\href}[2]{#2}\begingroup\raggedright\endgroup

\end{document}